\documentclass[aps,prl,showpacs,groupedaddress,twocolumn]{revtex4-1}
\usepackage{graphicx}
\usepackage{graphics}
\usepackage{amsmath,amssymb,amsfonts}
\usepackage{bm}
\begin{document}
\allowdisplaybreaks
\title{Giant circular dichroism of a molecule in a region of strong plasmon resonances between two neighboring gold nanocrystals}

\author {Hui Zhang,$^1$}
\altaffiliation{zhangh@ohio.edu}
\author {A. O. Govorov$^{1, }$}
\email{govorov@helios.phy.ohiou.edu}
\address{
$^1$Department of Physics and Astronomy, Ohio University, Athens, Ohio 45701, USA}
\begin{abstract}
We report on giant circular dichroism (CD) of a molecule inserted into a plasmonic hot spot. Naturally occurring molecules and biomolecules have typically CD signals in the UV range, whereas plasmonic nanocrystals exhibit strong plasmon resonances in the visible spectral interval. Therefore, excitations of chiral molecules and plasmon resonances are typically off-resonant. Nevertheless, we demonstrate theoretically that it is possible to create strongly-enhanced molecular CD utilizing the plasmons. This task is doubly challenging since it requires both creation and enhancement of the molecular CD in the visible region. We demonstrate this effect within the model which incorporates a chiral molecule and a plasmonic dimer. The associated mechanism of plasmonic CD comes from the Coulomb interaction which is greatly amplified in a plasmonic hot spot.   
\end{abstract} 

 \maketitle

\emph{Introduction.}
Plasmon resonances in metal nanostructures provide us with powerful methods of molecular spectroscopy. One impressive example is Surface Enhanced Raman Scattering (SERS) allowing the observation of vibrational spectra of single molecules \cite{Moskovits, Nie}. The physical mechanism of enhancement in SERS comes from amplification of electromagnetic fields in so-called plasmonic hot spots of metal nanocrystals. Another related example is an effect of plasmon-enhanced emission of dye molecules in the vicinity of metal nanoparticles (NPs) \cite{Lakowitz}. Circular dichroism (CD) is an optical technique very different from the methods mentioned above.  A CD spectrum is measured as a difference in absorbance of right-handed and left-handed circularly-polarized photons. Importantly, CD is sensitive to the symmetry of a molecule \cite{book}; more precisely, a CD spectrum tells us about the chiral property of a molecular object. Historically the CD method has been very successful in applications to biomolecules and drugs\cite{book}. For example, it was a powerful method to differentiate a secondary structure of proteins which can appear in the forms of $\alpha$-helix, random coil, or $\beta$-sheet. Recently the concepts of chirality and CD have been brought to the field of plasmonics \cite{Kuwata-Gonokami, Markovich, Thomas, Tang, Halas, Hendry, Gounko, Govorov-1, Fan, Slocik, Wang, Kitaev, Klimov, Hentschel}. One of the motivations for this development is plasmon-enhanced sensing of chiral molecules.

\begin{figure}[htbp]
\includegraphics[width=\columnwidth]{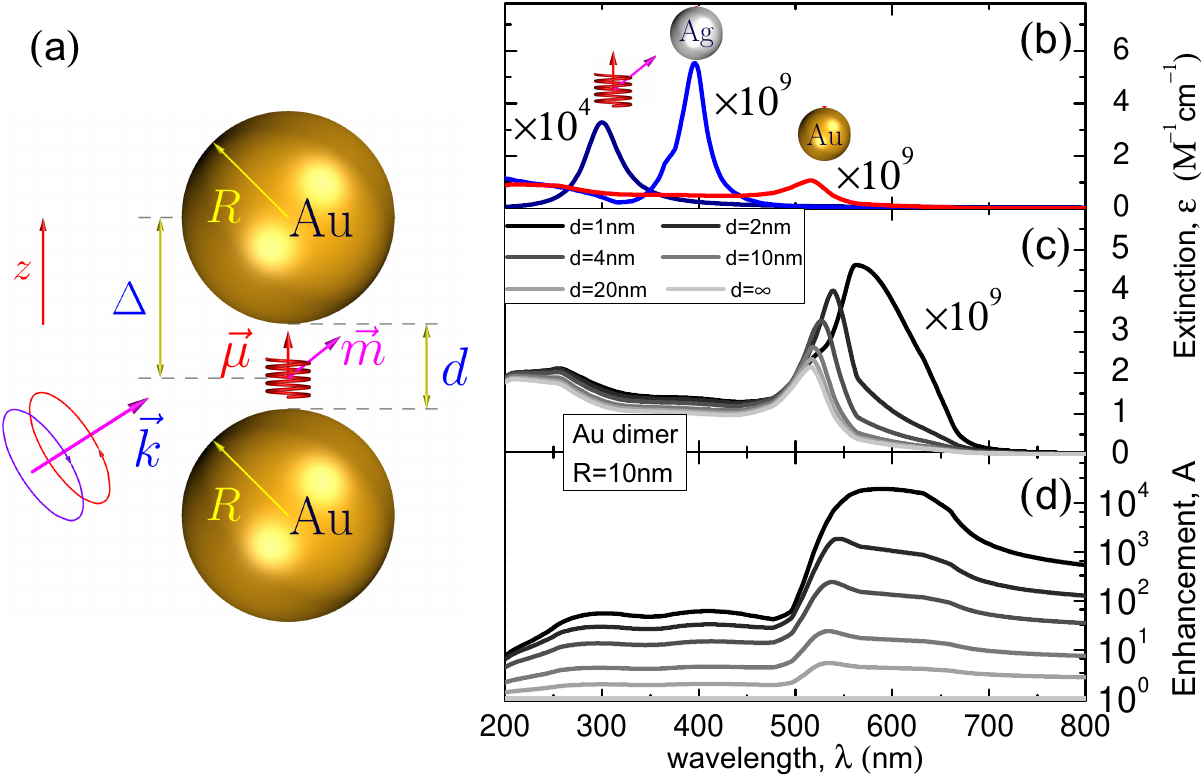}
\caption{ (color online) (a): Schematics of a model incorporating a gold dimer and a chiral molecule. (b): Extinctions of  single molecule and single noble metal NPs. (c): Extinction of Au dimer for various  separations $d$. (d): Enhancement factor $A$ versus the wavelength of incident light in the center of gold dimer in the absence of molecule. The parameters are: $\mathbf{E}_0||z$, molecular resonance $\lambda_0=300nm$.}\label{fig1}
\end{figure}

Here we show theoretically that the chiral property of a molecule can be both strongly enhanced and transferred to the visible wavelength range by using a plasmonic hot spot in a nanoparticle dimer [Fig.\ref{fig1}(a)].   For oriented molecules, the calculated enhancement factors of CD are 12 and 150 for the Au and Ag systems, respectively. For randomly-oriented chiral molecules, the enhancements are smaller, 4- and 12-fold. It is important that the enhancement effect is accomplished by the transfer of the CD signal from the UV to the visible-wavelength plasmonic resonances. Our results are in agreement with a recent experiment on nanocrystal aggregates and short DNA molecules \cite{Gounko}. In particular, this experimental paper \cite{Gounko} identified the aggregation of nanocrystals as a condition to observe enhanced plasmonic CD. Aggregates are expected to have hot plasmonic spots in gaps between nanocrystals. We also should note that our previous calculations \cite{Govorov-1, Govorov-2} have been performed mostly for single nanocrystals in the absence of plasmonic hot spots with giant enhancement and, as a result, we did not predict plasmon enhancement of CD, but we found an important effect of transfer of molecular chirality to the plasmon resonances; the plasmonic CD effect has been recently observed in experiments with Au-NPs and proteins \cite{Slocik}. As an important next step in the future developments, the plasmon enhancement effect can be extended to  the concept of Raman optical activity \cite{Barron}.  

\emph{Model.} A molecule-gold dimer hybrid system [Fig.\ref{fig1}(a)] is excited by an incident light wave with an electric field $\mathbf{E}_{\text{ext}}=\mathbf{E}_0e^{-i \omega t}+\mathbf{E}_0^\ast e^{i \omega t}$, where $\mathbf{E}_0$ is an amplitude; correspondingly, the magnetic field amplitude $\mathbf{B}_0=\frac{\sqrt{\epsilon_{r,0}}}{c}\mathbf{k}_0\times\mathbf{E}_0/k_0$. Here $\epsilon_{r,0}$ and $k_0=2\pi/\lambda$ are the relative dielectric constant of matrix (water) and the photon wavenumber in vacuum, respectively. We consider here the quasi-electrostatic case, i.e., $\lambda\gg D$, where $D$ is a size of the system. Then, the master equation for the quantum states of molecule reads:
\begin{equation}\label{e2}
\hbar \frac{\partial \rho_{ij}}{\partial t}=i[\hat{\rho},\hat{H}]_{ij}-\Gamma_{ij}(\rho), 
\end{equation}
where $\hat{H}=\hat{H}_0+\hat{H}^\prime$ is the Hamiltonian and $\hat{\rho}$ is the density matrix. Here $H_0$ describes the internal electronic structure of a molecule and $\hat{H}^\prime=-\hat{\bm \mu}\cdot\mathbf{E}_T-\hat{\mathbf{m}}\cdot\mathbf{B}_T+V_{\text{quad}}$ is the light-matter interaction operator; $\mathbf{E}_T$ and $\mathbf{B}_T$ are the fields acting on the molecule, and $\hat{\bm \mu}$ and $\hat{\mathbf{m}}$ stand for the electric and magnetic dipolar operators, respectively: $\hat{\bm \mu}=e\hat{\mathbf{r}},\ \hat{\mathbf{m}}=\frac{e}{2m_e}\hat{\mathbf{r}}\times\hat{\mathbf{p}}$; $V_{\text{quad}}$ is the quadrupole interaction  defined in the Supplementary Materials.
By using the rotating-wave approximation in the linear regime  and involving only two states (1 and 2) we obtain from  Eq.\eqref{e2} (see Supplementary Materials):
\begin{align}\label{e3}
&\rho_{21}=\sigma_{21}e^{-i \omega t}, \rho_{12}=\rho_{21}^\ast\nonumber, \mathbf{d}=\sigma_{21}\bm \mu_{12}, \\
&\sigma_{21}=-\frac{\bm \mu_{21}\cdot{\mathbf{E}^\prime}+\mathbf{m}_{21}\cdot \mathbf{B}^{\prime}}{\hbar(\omega-\omega_0)+i\gamma_{21}-G},\nonumber\\
&\rho_{22}=\frac{2\gamma_{21}}{\gamma_{22}}\frac{\left|\bm \mu_{21}\cdot\mathbf{E}^\prime+\mathbf{m}_{21}\cdot \mathbf{B}^{\prime}\right|^2}{\left|\hbar(\omega-\omega_0)+i\gamma_{21}-G\right|^2},
\end{align}
where $\hbar\omega_{1(2)}$ are the energies of the molecular states, $\omega_0=\omega_2-\omega_1$ is the frequency of molecular transition, and $\mathbf{E}^\prime$ is the electric field induced by the incident wave in the nanostructure at the position of a molecule. This electric field includes the external field and the plasmonic field induced by NPs. The magnetic field in a small structure is not affected much by the plasmonic effects, i.e.  $\mathbf{B}^\prime\approx \mathbf{B}_0$\cite{Govorov-1}. Besides, the broadening function $G=\frac{1}{4\pi \epsilon_0}\bm \mu_{21}\cdot\nabla(\bm \Phi_\omega\cdot \bm \mu_{12})+G_{\text{quad}}$, where the vector $\bm \Phi_\omega$ defines the electric potential $\Phi_d$ which is induced by the surface charges of NPs in the presence of an oscillating molecular dipole $\mathbf{d}$; then, the function $\varphi_d=\frac{1}{4\pi \epsilon_0}(\frac{\mathbf{d}\cdot\mathbf{r}}{r^3}+\Phi_d)$ gives the total electric potential induced by the molecular dipole in the presence of NPs, and $\Phi_d=\mathbf{d}\cdot \bm\Phi_\omega$. Thus, the total field acting on a molecule is given by $\mathbf{E}_T=\mathbf{E}^\prime-\frac{1}{4\pi \epsilon_0}\nabla  \Phi_d$. The function $G$ in Eqs.\eqref{e3} originated from the plasmonic surface charges induced by the molecular dipole  $\mathbf{d}$. Then, the total absorption rate of the system is
\begin{equation}\label{e4}
Q=Q_{\text{mol}}+Q_{\text{NP}},
\end{equation}
where $Q_{\text{NP}}=2 \omega \mathbf{Im}\epsilon_{\text{NP}} \int_{\text{NPs}}|\mathbf{ E}_{\text{tot}}|^2dV$ and $Q_{\text{mol}}=\omega_0 \rho_{22} \gamma_{22}$, in which $\epsilon_{\text{NP}}$ is the permittivity of NPs ($\epsilon_{NP}=\epsilon_{r,NP} \epsilon_{vac}$), the integral is taken over NP volumes, and $\mathbf{E}_{\text{tot}}$ denotes the total field inside NPs. A CD signal is defined as the difference between absorption of left- and right-handed polarized light, \begin{equation}\label{e5}
\text{CD}=\langle Q_+-Q_-\rangle_\Omega=\text{CD}_{\text{mol}}+\text{CD}_{\text{NP}},
\end{equation}
where $\langle\ldots\rangle_\Omega$ is the average over the solid angle of the direction of incident light; this averaging is equivalent to what happens in experiments performed in solution:  hybrid NP complexes have random orientations, but the incident light has a well-defined direction. Besides, $\pm$ denotes the two polarizations of incident field $\mathbf{e}_{0\pm}=(\mathbf{e}_\theta\pm i\mathbf{e}_\phi)/\sqrt{2}$. Since the total field inside NPs is $\mathbf{E}_T=\mathbf{E}^\prime-\nabla \varphi_d$, the important plasmonic contribution $Q_{\text{NP}}$ can be split into four terms $Q_{\text{NP}}=Q_0+Q_{\text{NP, diople-field}}+Q_{\text{NP, dipole-dipole}}+Q_{\text{NP, quad}}$, where $Q_0$ is the absorption by the NP dimer, 
\begin{equation}\label{add3}
\begin{aligned}
&Q_{\text{NP, dipole-field}}=-\frac{\omega}{\pi \epsilon_0} \mathbf{Im}[\epsilon_{\text{NP}}]\mathbf{Re}\int_{\text{NPs}} dV (\mathbf{E}^{\prime\ast}\cdot\nabla \varphi_{d}),\\
&Q_{\text{NP, dipole-dipole}}=\frac{\omega}{8\pi^2 \epsilon_0^2} \mathbf{Im}[\epsilon_{\text{NP}}]\int_{\text{NPs}} dV \left|\nabla \varphi_{d}\right|^2,
\end{aligned}
\end{equation}

Therefore, the $\text{CD}_{\text{NP}}$ should have three terms since the NP system itself is non-chiral and $Q_0$ does not give contribution to CD. Finally, we arrive to (see Supplementary Materials): 
\begin{align}\label{e6}
&\text{CD}_{\text{mol}}=\frac{8\sqrt{\epsilon_r}}{3c}\frac{\omega_0\gamma_{21}|E_0|^2\mathbf{Im}[\bm m_{21}\cdot(\hat{\mathcal{P}}^\dagger \bm \mu_{12})]}{\left|\hbar(\omega-\omega_0)+i\gamma_{21}-G\right|^2}+\text{CD}_{\text{mol, quad}},\nonumber\\
&\text{CD}_{\text{NP}}=\text{CD}_{\text{NP, dipole-field}}+\text{CD}_{\text{NP, dipole-dipole}}+\text{CD}_{\text{NP, quad-field}},\nonumber\\
&\text{CD}_{\text{NP, dipole-field}}=\frac{2\omega\sqrt{\epsilon_r}}{3c\pi\epsilon_0}|E_0|^2(\mathbf{Im} \epsilon_{\text{NP}})\nonumber\\
&\hspace{6em}\cdot\mathbf{Im}\int_{\text{NPs}} \frac{\bm m_{21}\cdot [\hat{K}^\dagger(\bm r)\nabla(\bm \Phi^{\text{tot}}(\bm r)\cdot \bm\mu_{21})]}{\hbar(\omega-\omega_0)+i \gamma_{21}-G} dV,\nonumber\\
&\text{CD}_{\text{NP, dipole-dipole}}=-\frac{8\omega\sqrt{\epsilon_r}}{3c}\frac{|E_0|^2\mathbf{Im}[\bm m_{21}\cdot(\hat{\mathcal{P}}^\dagger\bm \mu_{12})]}{|\hbar(\omega-\omega_0)+i \gamma_{21}-G|^2}\mathbf{Im}G,
\end{align}
where $\hat{\mathcal{P}}$ and $\hat{K}(\mathbf{r})$ are the field-enhancement matrices inside a molecule and NPs, respectively (see Supplementary Materials): $\mathbf{E}^\prime|_{\mathbf{r}=\mathbf{r}_{mol}}=\hat{\mathcal{P}}\mathbf{E}_0, \mathbf{E}^\prime(\mathbf{r})=\hat{K}(\mathbf{r})\mathbf{E}_0\ (\mathbf{r}\in \text{NPs})$. Besides, $\bm \Phi^{\text{tot}}$ is the total electric potential induced by dipole $\mathbf{d}$ inside NPs: $\bm \Phi^{\text{tot}}=\frac{\mathbf{r}}{r^3}+\bm \Phi_\omega$. 

In our study, we solve numerically the near-field Poisson equation within the local dielectric constant model and beyond the dipole limit (Supplemental Materials). We expand the electric potentials coming from the induced surface charges of NPs and the external field in terms of spherical harmonics (Supplementary Materials). For example, the potential induced by surface charges of one of the NPs can be written as $\bm \Phi_\omega^{ \text{out}}=\sum_{l,m}\frac{\mathbf{C}_{lm}}{r^{l+1}}Y_l^m(\theta, \phi)$, where $\mathbf{C}_{lm}=(C_{lm}^x, C_{lm}^y, C_{lm}^z)$ are coefficients found from the boundary conditions. This method allows us to compute metal nanostructures with very strong plasmonic enhancements [Fig.\ref{fig1}]. The convergence of the multipole expiation method was carefully checked; the number $L_{max}$ in the calculations was around $50$ for $d=1nm$ and $120$ for $d=0.5nm$.    

For the molecular dipole, we choose numbers typical for the experiments. We first define convenient parameters: $\mu_{12}=|e|r_{12}$ and $\bm \mu_{12}\cdot \mathbf{m}_{21}/\mu_{12}= i|e|r_0 \omega_0 r_{21}/2$. In our calculations, we use $r_{12}=2\AA, r_0=0.05\AA, \gamma_{12}=0.3eV$. The key chiral parameter $r_0$ determines a CD strength of a molecule \cite{book}: $\text{CD}_{\text{mol},0}\sim\mathbf{Im}(\bm\mu_{12}\cdot\mathbf{m}) \sim r_0$. The parameter $r_0$ is a small number since chirality of a molecule is usually weak. The above parameters of a chiral molecule yield typical numbers for the molecular extinction and CD. For the dielectric functions of metals (Au and Ag), the Palik's data \cite{Palik} were adopted; the optical dielectric constant of matrix (water) $\epsilon_{r,0}=1.8$. 

\emph{Results.} First, we briefly discuss optical properties of the constituting elements [Fig.\ref{fig1}(b)]. Molecules and bio-molecules usually have CD lines in the UV or near-UV range. In Fig.\ref{fig1}(b) we show an example of extinction spectrum of a molecule with a resonance at $300$nm. Its maximum is $\sim10^4M^{-1}cm^{-1}$ which is a typical value for molecular systems. Extinctions of noble metal NPs (Au and Ag) are much larger and have plasmon resonances in the visible ($520$nm and $400$nm). Moreover, plasmonic NP complexes demonstrate both red shifts of plasmon peaks and remarkable enhancement of incident electromagnetic fields [Fig.\ref{fig1}(c) and (d)]. Especially strong enhancement occurs in the plasmonic hot spots\cite{Abajo,Savasta}, i.e., in the center of the dimer. The calculated enhancement factor $A=|\mathbf{E}^\prime|^2/|\mathbf{E}_0|^2$ can be as large as $10^4$ for small separations $d$. 

\begin{figure}[htbp]
\includegraphics[width=\columnwidth]{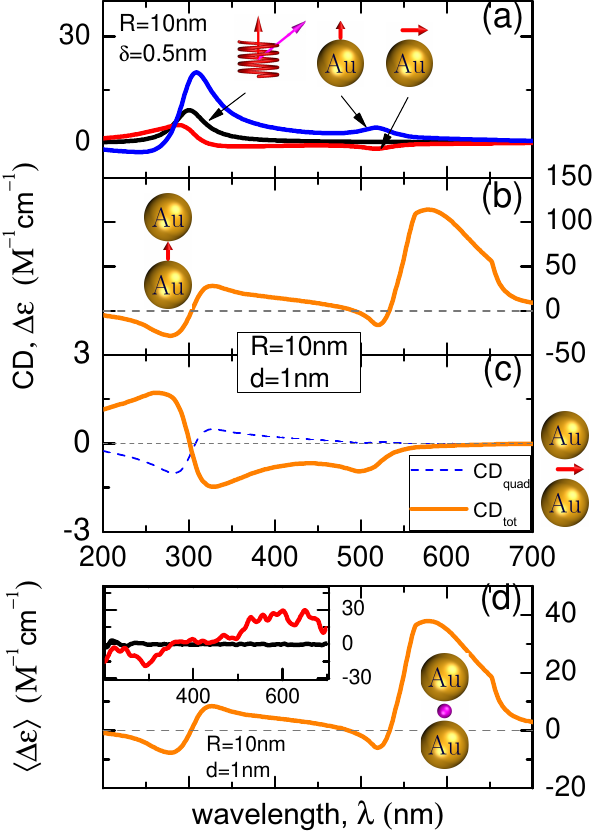}
\caption{ (color online) (a) CD spectra for an isolated molecule and for molecule-single NP complexes with two orientations of a molecular dipole: $\bm \mu || z$ and $\bm \mu || x$. (b) and (c): CD signals for the molecule-Au dimer complexes with different molecular orientations. (d) CD signal for the molecule-Au dimer complex averaged over the molecular dipole orientation; Inset shows experimental data obtained in \cite{Gounko}. The parameters of calculation are: $R=10nm, d=1nm, \lambda_0=300nm, \Delta=R+d/2$.}\label{fig2}
\end{figure}

\begin{figure}[htbp]
\includegraphics[width=\columnwidth, viewport=0 0 306 228, clip]{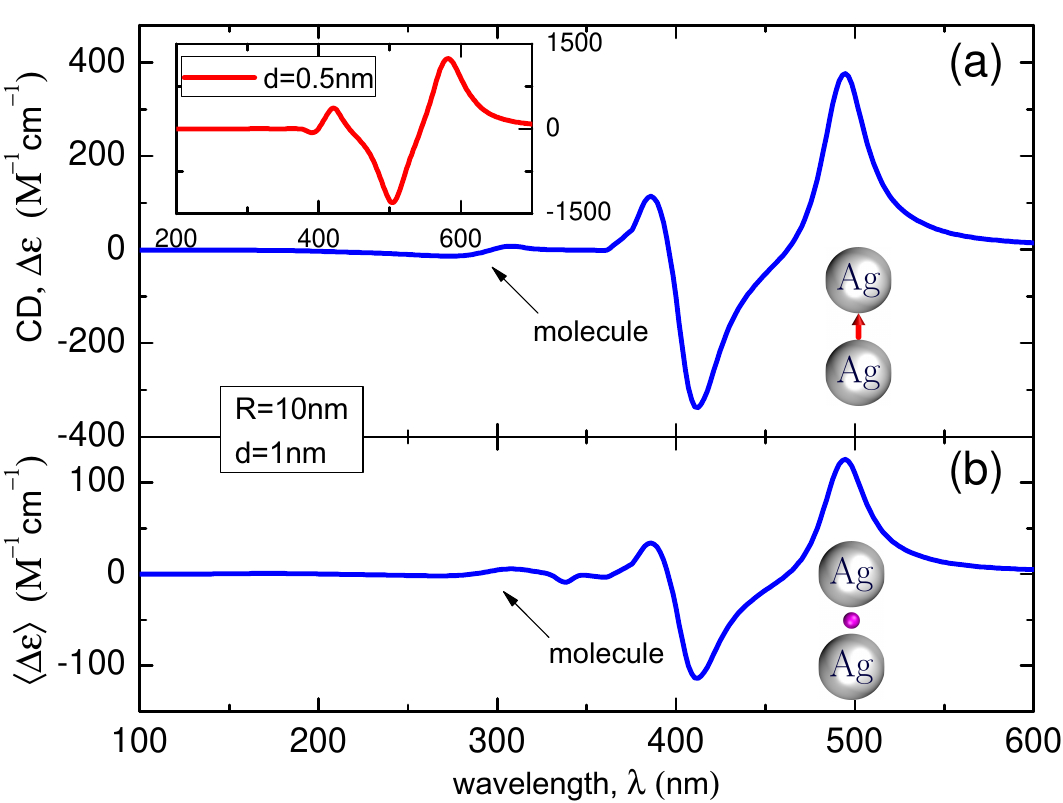}
\caption{ (color online) (a): CD spectrum of Ag dimer for $d=1$nm in the case $\bm \mu||z$. Inset shows the CD signal for the separation $d=0.5nm$. (b): CD spectrum of Ag dimer averaged over the molecular orientation. The parameters: $R=10nm$ and $\lambda_0=300nm$.}\label{fig3}
\end{figure}

Now we turn to the CD effect and first consider an off-resonant hybrid system with a gold dimer and a molecular resonance at $\lambda_0=300nm$ (Fig.\ref{fig2}). If a chiral molecule is placed in the vicinity of a single metal NP, the resultant CD spectrum acquires plasmon structures, but the strength of these structures is typically moderate \cite{Govorov-1,Slocik} [Fig.\ref{fig2}(a)]. A sign of the plasmonic CD depends on the orientation of molecular dipole. Remarkably different spectra appear for the chiral molecule in a metal NP dimer with small separations $d$ (Figs. \ref{fig2} and \ref{fig3}). For the Au dimer with ($\bm \mu_{12}|| z$),  [Fig.\ref{fig2}(b)], we obtain the following picture: The molecular CD band at $300nm$ is now transformed into an asymmetric structure and the plasmonic CD structure becomes very strong and broad. The first effect (i.e. the UV-CD structure) comes from the interference of incident and induced fields (Fano effect)\cite{Zhang, Neuhauser, Savasta}, whereas the second originates from the giant enhancement of electric field at the chiral molecule and from Coulomb interaction between a chiral object (molecule) and a non-chiral plasmonic nanostructure. We now introduce a figure of merit for the plasmon enhancement of CD: 
\begin{equation}\label{add4}
P_{\text{CD}}=\frac{\text{CD}_{max}(\lambda=\lambda_{plasm})}{\text{CD}_{mol,0,max}(\lambda=\lambda_{0})}. 
\end{equation}
This enhancement factor, which compares a CD signal in the plasmonic band with the original molecular CD at $300nm$, is $\sim12$. We also note that the plasmonic CD structure is broad because the inter-particle Coulomb interaction allows for excitation of many multipole modes of single NPs. For the other orientation ($\bm \mu_{12}\perp z$), the plasmonic CD band is weak because of the screening effect. The CD spectrum averaged over the orientation of the molecule should be calculated as: $\langle \text{CD}\rangle=(\text{CD}_z+\text{CD}_x+\text{CD}_y)/3\approx \text{CD}_z/3$. This spectrum still has a strong plasmon resonance (enhancement factor $\sim4$).  Remarkably, this enhancement is comparable with the experimental data \cite{Gounko} obtained for aggregated Au NPs and short DNA molecules [inset of Fig.\ref{fig2}(d)]. Besides, Fig.\ref{fig2}c also shows the contribution of the quadrupole interaction. Overall we found that the quadrople effect in a molecule is not enhanced by a plasmonic dimer and, due to the symmetry of the dimer,  can play a role only for the configurations $\bm \mu||x(y)$ (For details, see Supplementary Materials).

Fig.\ref{fig3} shows CD spectra for the Ag nanostructure which has much stronger plasmonic enhancement since the Ag plasmon resonance is much sharper. In this case, the configuration with $\bm \mu||z$ has a CD enhancement factor of $150$ for $d=0.5nm$ and an averaged CD is enhanced $12$-fold for $d=1nm$.   

\begin{figure}[htbp]
\includegraphics[width=\columnwidth, viewport=0 0 285 214, clip]{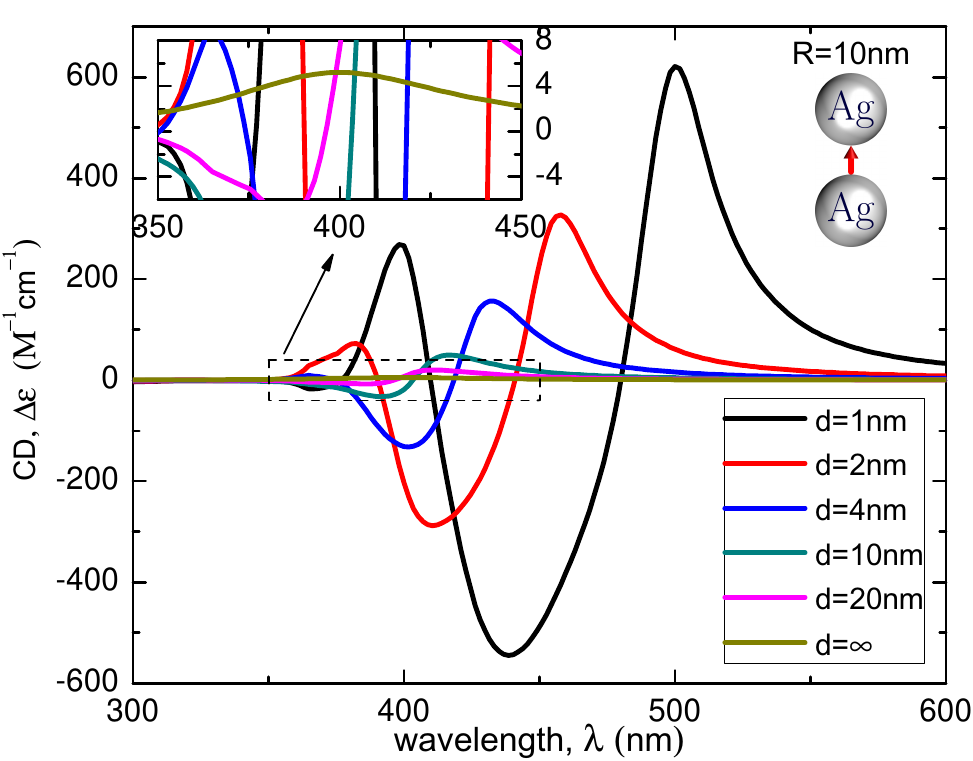}
\caption{ (color online) CD spectrum of Ag dimer for the molecular resonance $\lambda_0=400nm$ and for various separations $d$. Inset: Small region of the spectrum.}\label{fig4}
\end{figure}

Some of the biomolecules may have CD signals also in the blue spectral interval \cite{Nadia A. Abdulrahman}. This case is shown now in Fig.\ref{fig4}. A molecule with $\lambda_0=400nm$ is located in the center of an Ag dimer. The following effects should be now noticed: (a) The CD spectrum is amplified $125$-fold; (b) multipole plasmon resonances in the Ag-dimer create a broad CD spectrum with positive and negative bands; and (c) a strong plasmonic CD signal appears at $500nm$ which is a significantly longer wavelength compared with the original molecular CD wavelength of $400nm$. In other words, we again observe simultaneously a strong enhancement effect and a creation of the optical chirality in the red. Regarding the feature (b), we confirm the explanation for this effect by looking at the absorption spectrum of Ag dimer which indeed has strong multipole resonances in the interval $400-500nm$. Similar results can be obtained for the Au dimer.

We now look at the mechanism of the enhanced plasmonic CD. Numerical data shown in Supplementary Materials indicate that for the most common case (off-resonant exciton-plasmon regime), like shown in Figs. 2 and 3, the leading term in the CD signal is $\text{CD}_{\text{dipole-field}}$. This is the term coming from the interference of the incident field and the field induced by a chiral molecular dipole inside the NPs. In the off-resonant regime ($\omega_0-\omega_{plasm}\gg\gamma_{12}, |G|$), a CD signal in the plasmonic band is proportional the optical rotatory dispersion (ORD) of a molecule
\begin{equation}\label{add5}
\text{CD}_{\text{dipole-field}}\sim f(\omega)\cdot\text{ORD}_{\text{mol,0}}\sim f(\omega)\frac{m_{12}\mu_{21}}{\omega_0-\omega_{\text{plasm}}},
\end{equation}
where $f(\omega)$ is a complex function describing the energy dissipation inside the metal in the vicinity of molecule placed in a plasmonic hot spot. Since the function $f(\omega)$ comes from the induced dissipative currents inside the plasmonic NPs, we expect that this function should be greatly amplified at the plasmon resonances because of the enhancement of the electric fields in a hot spot.  Indeed we see this effect in our data (Figs.\ref{fig2} and \ref{fig3}). In our physical model, the enhanced CD originates from a joint action of the two key effects: the ORD response of a chiral molecule at the plasmon wavelength and the plasmonic hot spot effect. We should note here that the molecule itself does not absorb at the plasmon resonance. Therefore, the giant CD at the plasmon wavelength comes from the dissipation inside the metal subsystem.  The plasmon resonances of non-chiral metal NPs interact with a chiral non-absorbing environment (a chiral molecule in the hot spot) and therefore acquire a chiral property, and demonstrate CD.  Another qualitative observation is that the plasmonic CD decreases with $\omega_0-\omega_{\text{plasm}}$ since it depends on the molecular ORD at the plasmonic frequency (Eq. 8).

To conclude, we have presented a theory of optical activity of a chiral molecule inserted into a plasmonic hot spot and we have found an important effect - simultaneous strong plasmonic enhancement and shift of optical chirality from the UV range to the visible. The mechanism of enhanced plasmonic CD signals is based on the Coulomb interaction between a chiral molecules and plasmonic modes in a hot spot. Importantly, single spherical NPs are not able to create any significant enhancement of CD signals and, therefore, the use of plasmonic hot spots is crucial for the ultra-sensitive CD spectroscopy. The effect of strong plasmonic CD in the visible, which is described here, can be utilized for chiral bio-sensing and also for construction of novel chiral optically-active materials.  

\emph{Acknowledgment.} This work was supported by Volkswagen Foundation and NSF (Project: CBET-0933415).


\end{document}


\allowdisplaybreaks
\section{Absorption rate of chiral molecule}
Next we shall derive the formula Eq.(2) in the Lett.

The master equation for the quantum states of molecule is given by:
\begin{equation}\label{s1}
\hbar \frac{\partial \rho_{ij}}{\partial t}=i[\rho,\hat{H}]-\Gamma_{ij}(\rho)
\end{equation}
in which $\hat{H}=\hat{H}_0+\hat{H}^\prime$ is the Hamiltonian, and  $\hat{H}^\prime=(-\hat{\bm \mu}\cdot\mathbf{E}_T-\hat{\mathbf{m}}\cdot\mathbf{B}_T)$ where the total field are respectively $\mathbf{E}_T=\mathbf{E}_{\text{tot}}e^{-i \omega t}+\mathbf{E}_{\text{tot}}^\ast e^{i \omega t}, \mathbf{B}_T=\mathbf{B}_{\text{tot}}e^{-i \omega t}+\mathbf{B}_{\text{tot}}^\ast e^{i \omega t}$. Here, $\mathbf{E}_T, \mathbf{B}_T$ are the total electromagnetic fields inside chiral molecule. Besides, $\bm \mu=e\mathbf{r},\ \mathbf{m}=\frac{e}{2m_e}\mathbf{r}\times\mathbf{p}$ stand for the electric and magnetic dipolar operators, respectively. $\Gamma_{ij}(\rho)$ is the relaxation term, and if we only involve two quantum states, it will be given by: $\Gamma_{11}(\rho)=-\Gamma_{22}(\rho)=- \gamma_{22} \rho_{22}, \Gamma_{12}(\rho)=\gamma_{21}\rho_{12}, \Gamma_{21}(\rho)=\gamma_{21}\rho_{21}$. Here we note that the effects of the quadrupole light-matter interaction will be discussed separately in the last section of this Supplementary Material.  Then, after substituting $H$ into \eqref{s1}, one has:
\begin{align}
&\hbar\frac{\partial \rho_{21}}{\partial t}=-i\hbar \omega_0 \rho_{21}+i(\rho_{11}-\rho_{22})({\bm \mu}_{21}\cdot\mathbf{E}_T+\mathbf{m}_{21}\cdot\mathbf{B}_T)-
\gamma_{21}\rho_{21}\label{s2}\\
&\hbar\frac{\partial \rho_{22}}{\partial t}=-i(\rho_{21}{\bm \mu}_{12}-{\bm \mu}_{21}\rho_{12})\cdot\mathbf{E}_T-i(\rho_{21}\mathbf{m}_{12}-\mathbf{m}_{21}\rho_{12})\cdot\mathbf{B}_T-\gamma_{22}
\rho_{22}\label{s3}
\end{align}
where $\bm \mu_{i,j}=\langle i|\hat{\bm \mu}|j\rangle, i=1,2$, and $\omega_0=\omega_2-\omega_1$. Next, we adopt the rotating-wave approximation $\rho_{21}=\sigma_{21}e^{-i \omega t}, \rho_{12}=\rho_{21}^\ast$ and only keep the $e^{- i \omega  t}$ time dependence in Eqs. \eqref{s2} and \eqref{s3}, thus Eqs. \eqref{s2} and \eqref{s3} become: 
\begin{align}
&\hbar\frac{\partial \sigma_{21}}{\partial t}\approx\left[i\hbar(\omega-\omega_0)-\gamma_{21}\right] \sigma_{21}+i(\rho_{11}-\rho_{22})({\bm \mu}_{21}\cdot\mathbf{E}_{\text{tot}}+\mathbf{m}_{21}\cdot\mathbf{B}_{\text{tot}})
\label{s4}\\
&\hbar\frac{\partial \rho_{22}}{\partial t}\approx -i\left[\sigma_{21}({\bm \mu}_{12}\cdot\mathbf{E}_{\text{tot}}^\ast+\mathbf{m}_{12}\cdot
\mathbf{B}_{\text{tot}}^\ast)-\sigma_{12}({\bm \mu}_{21}\cdot\mathbf{E_{\text{tot}}}+\mathbf{m}_{21}\cdot
\mathbf{B}_{\text{tot}})\right]-\gamma_{22}\rho_{22}\label{s5}
\end{align}
Since $\sigma_{21}$ is a new `slowly' varying variable, thus we have $\hbar\frac{\partial \sigma_{21}}{\partial t}\approx 0$ and $\hbar\frac{\partial \rho_{22}}{\partial t}\approx 0$. 
Furthermore, $\mathbf{E}_{\text{tot}}$ inside molecule can be divided into two parts: $\mathbf{E}_{\text{tot}}=\mathbf{E}^\prime+\mathbf{E}_d, \mathbf{E}^\prime\equiv\mathbf{E}_0+\mathbf{E}_{ind}$, where $\mathbf{E}_0$ is the incident field, $\mathbf{E}_{ind}$ is the electric field induced by NPs in the absence of chiral molecule, and $\mathbf{E}_d$ is electric field coming from charges in surface of NPs induced only by molecule. Noticeable, in the presence of electromagnetic filed, a molecular dipole moment $\mathbf{d}_{\text{mol}}=\mathbf{Tr}(\rho\bm \mu)=\mathbf{d} e^{-i \omega t}+\mathbf{d}^\ast e^{i \omega t}$ will be induced where $\mathbf{d}=\sigma_{21}\bm \mu_{12}$, and accordingly $\mathbf{E}_d$ has the following form:
\begin{equation}\label{s6}
\mathbf{E}_d=-\frac{1}{4\pi \epsilon_0}\nabla(\bm \Phi_\omega\cdot \mathbf{d})=-\frac{1}{4\pi \epsilon_0}\nabla(\bm \Phi_\omega\cdot \bm \mu_{12}) \sigma_{21}
\end{equation}
where $\epsilon_0$ is the electric permittivity of space, and we adopt the Palik data hereafter. The absorption rates of single Au (Ag) sphere are shown in Fig.1(b) of the Lett.  Therefore, Eq.\eqref{s4} gives us:
\begin{align}\label{s6}
&0\approx\left[i\hbar(\omega-\omega_0)-\gamma_{21}\right] \sigma_{21}+i(\rho_{11}-\rho_{22})\left[{\bm \mu}_{21}\cdot\left(\mathbf{E}^\prime-\frac{1}{4\pi \epsilon_0}\nabla(\bm \Phi_\omega\cdot \bm \mu_{12}) \sigma_{21}\right)+\mathbf{m}_{21}\cdot\mathbf{B}^{\prime}\right]\nonumber\\
&\Rightarrow \sigma_{21}=-\frac{(\rho_{11}-\rho_{22})(\bm \mu_{21}\cdot{\mathbf{E}^\prime}+\mathbf{m}_{21}\cdot \mathbf{B}^{\prime})}{\hbar(\omega-\omega_0)+i\gamma_{21}-(\rho_{11}-\rho_{22})G}
\end{align}
where the $G$ function is defined as $G=\frac{1}{4\pi \epsilon_0}\bm \mu_{21}\cdot\nabla(\bm \Phi_\omega\cdot \bm \mu_{12})$ and we have used $\mathbf{B}_{\text{tot}}\approx \mathbf{B}^\prime\approx \mathbf{B}_0$. Similarly, from \eqref{s5} we have:
\begin{equation}\label{s7}
\rho_{22}=\frac{2\gamma_{21}}{\gamma_{22}}\frac{(\rho_{11}-\rho_{22})\left|\bm \mu_{21}\cdot\mathbf{E}^\prime+\mathbf{m}_{21}\cdot \mathbf{B}^{\prime}\right|^2}{\left|\hbar(\omega-\omega_0)+i\gamma_{21}-(\rho_{11}-\rho_{22})G\right|^2}
\end{equation}

In the linear regime $\rho_{22}\ll 1$, Eqs. \eqref{s6} and \eqref{s7} reduce to the formula Eq.(2) in the Lett. Finally, by using $Q_{\text{mol}}=\omega_0 \rho_{22} \gamma_{22}$, the absorption rate of chiral molecule in the linear regime is given by:
\begin{equation}\label{s8}
Q_{\text{mol}}=2 \omega_0 \gamma_{21}\frac{\left|\bm \mu_{21}\cdot\mathbf{E}^\prime+\mathbf{m}_{21}\cdot \mathbf{B}^{\prime}\right|^2}{\left|\hbar(\omega-\omega_0)+i\gamma_{21}-G\right|^2}
\end{equation}
\section{General dielectric formalism}
\begin{figure}[htbp]
\includegraphics[width=.9\columnwidth]{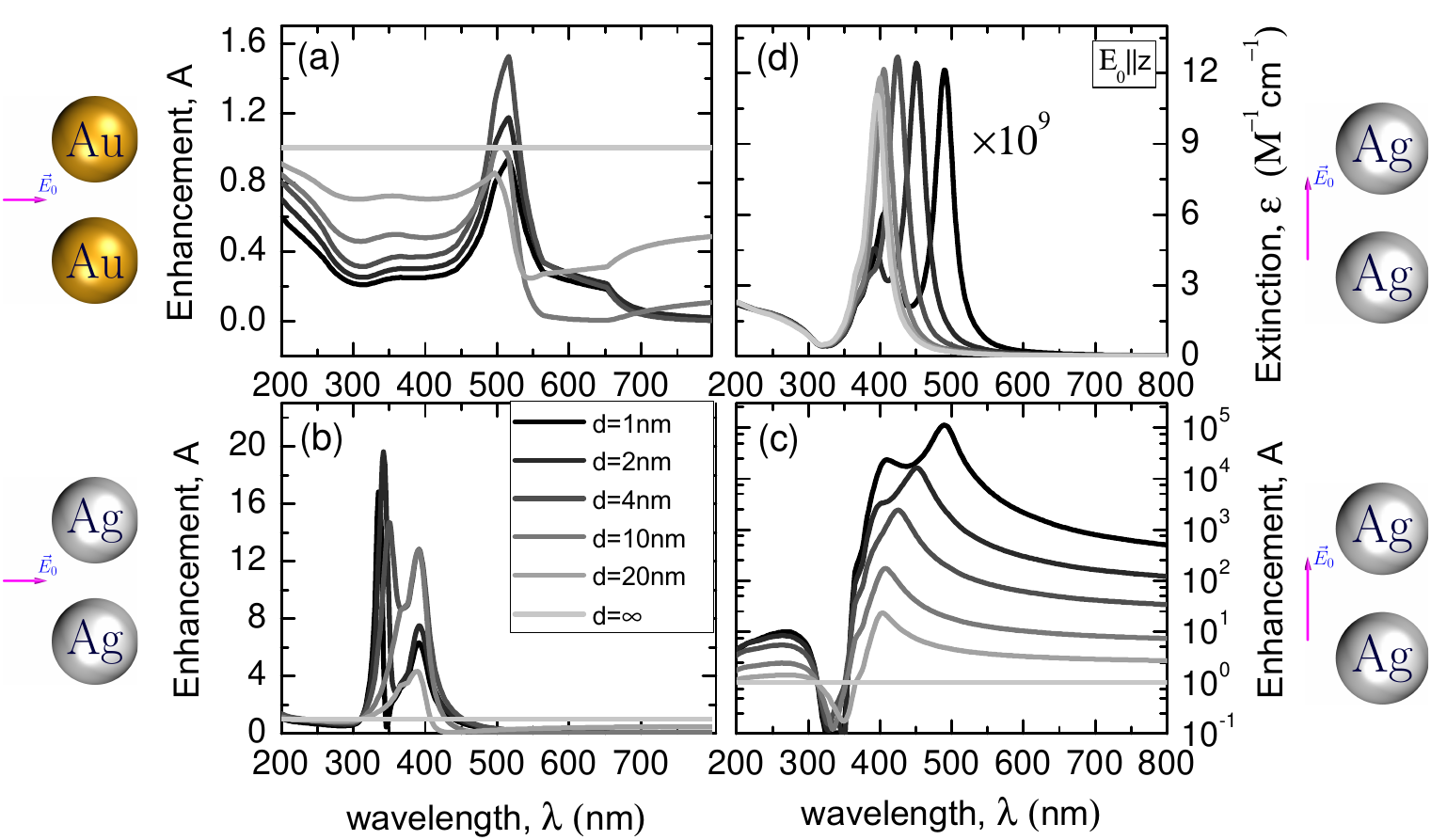}
\caption{ (color online) (a) shows enhancement factor $A$ versus the wavelength of incident light in the hot spot of Au dimer under the incident light $\mathbf{E}_0||x$, and (b) and (c) shows, respectively, the enhancement factor $A$ for the Ag dimer in the hot spot in the case $\mathbf{E}_0||x$ and $\mathbf{E}_0||z$. (d): Extinction of Ag dimer for various separations $d$ under $\mathbf{E}_0||z$. Other unmentioned parameters are the same as Fig.1 in the main text.}\label{SI7}
\end{figure}

The general dielectric formalism of molecule-NP dimer complex reads as follows:
\begin{equation}\label{s35}
\nabla\cdot \epsilon \nabla \phi_{ind}=\nabla\cdot \epsilon\mathbf{E}_0,\ \nabla\cdot \epsilon \nabla \varphi_d=\rho_d
\end{equation}
where $\epsilon$ denotes the dielectric function of medium and NPs, $\phi_{ind}$ is the  electric potential induced by NPs in the absence of molecule, $\mathbf{E}_0$ is the incident field, $\varphi_d$ is the electric potential of complex only induced by molecular dipole $\mathbf{d}$ in the absence of incident field, while $\rho_d$ is the charges inside molecule. Then, in the absence of molecule, the total electric field is given by: $\mathbf{E}^\prime=-\nabla \phi_{ind}+\mathbf{E}_0$, and the enhancement matrices can be defined as:
\begin{equation}\label{s36}
\mathbf{E}^\prime|_{\mathbf{r}=\mathbf{r}_{mol}}=\hat{\mathcal{P}}\mathbf{E}_0;\hspace{5em} \mathbf{E}^\prime(\mathbf{r})=\hat{K}\mathbf{E}_0\hspace{1em}(\mathbf{r}\in\text{NPs})
\end{equation}
$\hat{\mathcal{P}}$ and $\hat{K}$ are very useful for the following derivations. Similarly, from the second equation of \eqref{s35}, we can obtain the G function as well as electric potential induced by the molecular dipole $\varphi_d$. In Fig.\ref{SI7} we show the enhancement factor $A=|\mathcal{P}|^2$ in the hot spot of Au(Ag) dimer respectively.

\subsection{Electric potential induced only by molecular dipole}
 
First let us consider the isolated molecule case. Supposing the molecule is placed in the position $\mathbf{r}_d=(r_d, \theta_d, \phi_d)$ with $\theta_d=0 (\pi), \phi_d=0$, and the molecular dipole moment is $\mathbf{d}$, then the electric potential in the position $\mathbf{r}=(r, \theta, \phi)$ induced by this dipole moment is given by:
\begin{equation}\label{s9}
\varphi_0=\frac{1}{4\pi \epsilon_0}\mathbf{d}\cdot\bm \Phi_0, \bm \Phi_0=\sum_{lm}\begin{bmatrix}
B_{lm}^x\\B_{lm}^y\\B_{lm}^z
\end{bmatrix}\left(\frac{4\pi}{2l+1}\right)Y_l^m(\theta, \phi)
\end{equation}
with the coefficients given by:
\begin{align}
&B_{lm}^x=\pm\frac{1}{r_d}\left.\frac{\partial}{\partial \theta^\prime}\left[\frac{r_<^l}{r_>^{l+1}}Y_l^{m\ast}(\theta^\prime, \phi^\prime)\right]\right|_{\theta^\prime=0(\pi), \phi^\prime=0}= (-1)^{(l-1)(1-\alpha)/2}\frac{N_l^m}{r_d}\frac{r_<^l}{r_>^{l+1}}\left\{\begin{aligned}
&\frac{1}{2}&(m=-1)\\
&-\frac{l(l+1)}{2}&(m=1)\\
&0&(m\neq\pm1)
\end{aligned}\right.\label{s10}\\
&B_{lm}^y=\frac{1}{r_d\sin \theta^\prime}\left.\frac{\partial}{\partial \phi^\prime}\left[\frac{r_<^l}{r_>^{l+1}}Y_l^{m\ast}(\theta^\prime, \phi^\prime)\right]\right|_{\theta^\prime=0(\pi), \phi^\prime=0}=-imB_{lm}^x\label{s11}\\
&B_{lm}^z=\pm\left.\frac{\partial}{\partial r}\left[\frac{r_<^l}{r_>^{l+1}}Y_l^{m\ast}(\theta^\prime, \phi^\prime)\right]\right|_{r=r_d, \theta^\prime=0(\pi), \phi^\prime=0}=(-1)^{(l-1)(1-\alpha)/2}\delta_{m,0} N_l^m\begin{cases}
l\frac{r_d^{l-1}}{r^{l+1}}\hspace{2em}&(r_d<r)\\
-(l+1)\frac{r^l}{r_d^{l+2}}\hspace{2em}&(r_d>r)\label{s12}
\end{cases}
\end{align}
where $N_l^m=\sqrt{\frac{2l+1}{4\pi}\frac{(l-m)!}{(l+m)!}}, r_>=\max(r_d,r), r_<=\min(r_d,r)$, and $\alpha=\pm$ corresponds to $\theta_d=0, \pi$.

Next, we investigate the electric potential induced by dipole moment in the presence of nanoparticle (NP) dimer, and denote the coordinates system as $S$ ($S^\prime$) with the center of 1st (2nd) NP being the origin. We assume the electric potential coming from charges on surface of 1st and 2nd NP are [In coordinates $S$ ($S^\prime$) respectively]:
\begin{align}
&\varphi_1=\frac{1}{4\pi \epsilon_0}\mathbf{d}\cdot \bm \Phi_1&& \varphi_2=\frac{1}{4\pi \epsilon_0}\mathbf{d}\cdot \bm \Phi_2\label{s13}\\
&\bm \Phi_1=\left\{\begin{aligned}
&\sum_{lm}\begin{bmatrix}
C_{lm}^x\\C_{lm}^y\\C_{lm}^z
\end{bmatrix}\frac{r^l}{R^{2l+1}}Y_l^m(\theta, \phi)&(r<R)\\
&\sum_{lm}\begin{bmatrix}
C_{lm}^x\\C_{lm}^y\\C_{lm}^z
\end{bmatrix}\frac{1}{r^{l+1}}Y_l^m(\theta, \phi)&(r>R)
\end{aligned}\right.
&&\bm \Phi_2=\left\{
\begin{aligned}
&\sum_{lm}\begin{bmatrix}
D_{lm}^x\\D_{lm}^y\\D_{lm}^z
\end{bmatrix}\frac{{r^\prime}^l}{R^{2l+1}}Y_l^m( \theta^\prime, \phi^\prime)&(r^\prime<R)\\
&\sum_{lm}\begin{bmatrix}
D_{lm}^x\\D_{lm}^y\\D_{lm}^z
\end{bmatrix}\frac{1}{{r^\prime}^{l+1}}Y_l^m(\theta^\prime, \phi^\prime)&(r^\prime>R)
\end{aligned}\right.\label{s14}
\end{align}

Then, the field inside and outside 1st NP are given by (in coordinates $S$):
\begin{equation}\label{s15}
\psi_1=\frac{1}{4\pi \epsilon_0}\mathbf{d}\cdot \left\{\begin{aligned}
&\bm \Phi_0^{(1)}+\bm \Phi_1^{\text{in}}+\bm \Phi_2^{\text{out}}&(r<R)\\
&\bm \Phi_0^{(1)}+\bm \Phi_1^{\text{out}}+\bm \Phi_2^{\text{out}}&(r>R)
\end{aligned}\right.
\end{equation}
here $\bm \Phi_0^{(1)}$ is the potential induced by isolated molecule in coordinates $S$, i.e., $\alpha=-1$ and $\mathbf{r}_d$ is the position of molecule with respect to the center of 1st NP. Similar results of field inside and outside 2nd NP can also be obtained. Then, by using the boundary condition on each NP: 
\begin{equation}\label{s16}
\psi_i|_{r=R-0^+}=\psi_i|_{r=R+0^+}, \epsilon_{\text{NP}} \partial \psi_i/\partial r|_{r=R-0^+}=\epsilon_0 \partial \psi_i/\partial r|_{r=R+0^+}
\end{equation}
where $i=1,2$ and $\epsilon_{\text{NP}}$ is the dielectric function of NPs, the coefficients $C_{lm}^j, D_{lm}^j$ of $\bm \Phi_1, \bm \Phi_2$ can be obtained by solving a set of linear equations.  

\subsection{G function}
Once $\bm \Phi_1, \bm \Phi_2$ in Eq.\eqref{s14} are obtained, the dipolar part of $G$ function $G=\frac{1}{4\pi \epsilon_0}\bm \mu_{21}\cdot\nabla[(\bm \Phi_1^{\text{out}}+\bm \Phi_2^{\text{out}})\cdot \bm \mu_{12}]$ can be obtained as well through simple algebra. The final result is:
\begin{equation}\label{s17}
G=\frac{1}{4\pi \epsilon_0}\sum_{lm}\bm \mu_{21}\cdot\left\{\begin{bmatrix}
A_{lm}^x\\-A_{lm}^y\\A_{lm}^z
\end{bmatrix}\frac{C_{lm}^x \mu_{12x}+C_{lm}^y \mu_{12y}+C_{lm}^z \mu_{12z}}{{r_{d,1}}^{l+2}}+
\begin{bmatrix}
F_{lm}^x\\-F_{lm}^y\\F_{lm}^z
\end{bmatrix}\frac{D_{lm}^x \mu_{12x}+D_{lm}^y \mu_{12y}+D_{lm}^z \mu_{12z}}{{r_{d,2}}^{l+2}}\right\}
\end{equation}
where coefficients $C_{lm}^i, D_{lm}^i$ are the same as in Eq.\eqref{s14}, $r_{d,i}  (i=1, 2)$ is the distance of molecule with respect to the center of $i$th NP, and $A_{lm}^i (F_{lm}^i)$ is the same as coefficients $B_{lm}^i$ under $\alpha=-1 (\alpha=1)$ in Eq.\eqref{s10}$\sim$Eq.\eqref{s12} but the $r$ factor should be removed. Specifically, most $A_{lm}^i, F_{lm}^i$ are zeros expect for:
\begin{align}
&A_{lm}^x=(-1)^lN_l^m\begin{cases}
-\frac{1}{2}&(m=-1)\\
\frac{l(l+1)}{2}&(m=1)
\end{cases},\ A_{lm}^y=-im A_{lm}^x,\ A_{lm}^z=(-1)^lN_l^m(l+1)\ (m=0)\label{s18}\\
&F_{lm}^x=-N_l^m \begin{cases}
-\frac{1}{2}&(m=-1)\\
\frac{l(l+1)}{2}&(m=1)
\end{cases},\ F_{lm}^y=-imF_{lm}^x,\ F_{lm}^z=-N_l^m(l+1)\ (m=0)\label{s19}
\end{align}
where $N_l^m=\sqrt{\frac{2l+1}{4\pi}\frac{(l-m)!}{(l+m)!}}$.

\section{CD signal of molecule-NP dimer complex}

Next, we shall derive the formule Eq.(6) in the Lett.

The CD signal refers the the difference between absorption of left- and right-handed polarized light, so that it is given by:
\begin{equation}\label{s21}
\text{CD}=\langle Q_+-Q_-\rangle_\Omega=\text{CD}_{\text{mol}}+\text{CD}_{\text{NP}}
\end{equation}
where $\langle\ldots\rangle_\Omega$ is the average over solid angle of incident field which is equivalent to what happens in experiments: hybrid complex has random orientations but the incident light has well-defined direction. Besides, $\pm$ denotes the two polarizations of incident field $\mathbf{e}_{0\pm}=(\mathbf{e}_\theta\pm i\mathbf{e}_\phi)/\sqrt{2}$. In the following we shall calculate the two parts, i.e. $\text{CD}_{\text{mol}}$ and $\text{CD}_{\text{NP}}$ respectively.

\subsection{CD signal of molecular part}
Substituting the two polarizations into Eq.\eqref{s8}, we have:
\begin{align}\label{s22}
&Q_{\text{mol}\pm}=\omega_0\gamma_{21}\frac{|E_0|^2\left|\bm \mu_{21}\cdot \hat{\mathcal{P}} (\mathbf{e}_\theta\pm i\mathbf{e}_\phi)+\bm m_{21}\cdot \sqrt{\epsilon_r} (\mathbf{e}_\phi\mp i\mathbf{e}_\theta)/c\right|^2}{\left|\hbar(\omega-\omega_0)+i\gamma_{21}-G\right|^2}\nonumber\\
&\Rightarrow \text{CD}_{\text{mol}}=\langle Q_{\text{mol}+}-Q_{\text{mol}-}\rangle_\Omega=\frac{8\sqrt{\epsilon_r}}{3c}\frac{\omega_0\gamma_{21}|E_0|^2\mathbf{Im}[\bm m_{21}\cdot(\hat{\mathcal{P}}^\dagger \bm \mu_{12})]}{\left|\hbar(\omega-\omega_0)+i\gamma_{21}-G\right|^2}
\end{align}
where the enhancement matrix $\hat{\mathcal{P}}$ is defined in Eq.\eqref{s36}, and used the formula: $\mathbf{B}^\prime\approx \mathbf{B}_0$, \\$\langle\left|\bm \mu\cdot (\mathbf{e}_\theta\pm i\mathbf{e}_\phi)+\bm m\cdot(\mathbf{e}_\phi\mp i\mathbf{e}_\theta)\right|^2\rangle_\Omega=\frac{2}{3}|\bm \mu\mp i \bm m|^2$. Noticeable, $\mathbf{E}^\prime$ is the actual electric field in the center of NP dimer in the absence of molecule.

Once the CD signal Eq.\eqref{s22} is obtained, we can calculate it in the standard unit of $M^{-1}cm^{-1}$ as follows. Since the light intensity in this case is: $\langle S\rangle_t=2\epsilon_{vac}c\sqrt{ \epsilon_r}|\bm E_0|^2$, thus the cross section is (in unit of $m^2$):
\begin{align}\label{s23}
\Delta\sigma_{\text{mol}}\equiv\frac{\text{CD}}{\langle S\rangle_t}=\frac{4}{3\epsilon_{vac} c^2}\frac{\omega_0\gamma_{21}\mathbf{Im}[\bm m_{21}\cdot(\hat{\mathcal{P}}^\dagger \bm \mu_{12})]}{\left|(\omega-\omega_0)+i\gamma_{21}-G\right|^2}
\end{align}
where $\epsilon_{vac}$ here is the vacuum permittivity, $\epsilon_r=\epsilon_0/\epsilon_{vac}$, $c$ is speed of light in vacuum. Accordingly the CD signal in unit of $M^{-1}cm^{-1}$ is given by: \begin{equation}
\text{CD}_{\text{mol}}: \Delta \varepsilon_{\text{mol}}=\frac{N_A}{0.23}\Delta \sigma_{\text{mol}}
\end{equation}
in which $N_A$ is the Avogadro constant. 

\subsection{CD signal of NP part}

The absorption rate of NPs is given by:
\begin{equation}\label{s24}
Q_{\text{NP}}(\omega)=\sum_iQ_i(\omega)=2 \omega\sum_i(\mathbf{Im}\epsilon_{\text{NP}})\int|{\bm E_i}_{\text{tot}}|^2dV
\end{equation}
where $\epsilon_{\text{NP}}$ is the dielectric function of NPs, ${\bm E_i}_{\text{tot}}$ is the 
total electric field inside the $i$th sphere: $\bm E_{i\text{tot}}=\bm E^\prime-\frac{1}{4\pi \epsilon}\nabla \phi_{\text{dipole}}=\bm E^\prime-\frac{1}{4\pi \epsilon}\nabla(\bm \Phi^{\text{tot}}\cdot \bm \mu_{12})\sigma_{21}$ in which $\epsilon_0$ is the electric permittivity of space, and $\bm \Phi^{i,\text{tot}}(\bm r)=\bm \Phi_0+\bm \Phi_1+\bm \Phi_2$ is the total electric potential in $i$th sphere: $\bm r\in i$th sphere. Considering that $\text{CD}_{\text{NP}}=\langle Q_{\text{NP},+}-Q_{{\text{NP}},-}\rangle_\Omega$, thus we have:
\begin{align}
&\text{CD}_{\text{NP}}=\text{CD}_{\text{NP, dipole-field}}+\text{CD}_{\text{NP, dipole-dipole}}\label{s25}\\
&\text{CD}_{\text{NP, dipole-field}}=-\frac{\omega}{\pi \epsilon_0}\sum_i(\mathbf{Im}\epsilon_{\text{NP}})\mathbf{Re}\left\langle\int\left[\left(\bm E_i^{\prime\ast}\cdot\nabla \phi_{\text{dipole},i}\right)_+-\left(\bm E_i^{\prime\ast}\cdot\nabla \phi_{\text{dipole},i}\right)_-\right]dV\right\rangle_\Omega\label{s26}\\
&\text{CD}_{\text{NP, dipole-dipole}}=\frac{\omega}{8\pi^2 \epsilon_0^2}\sum_i(\mathbf{Im} \epsilon_{\text{NP}})\left\langle\int\left[{\left|\nabla \phi_{\text{dipole},i}\right|^2}_+-{\left|\nabla \phi_{\text{dipole},i}\right|^2}_-\right]dV\right\rangle_\Omega\label{s27}
\end{align}
where $\pm$ stands for the different incident direction $\bm e_{0\pm}=(\bm e_\theta\pm i\bm e_\phi)/\sqrt{2}$, $\epsilon_0$ is the permittivity of space, and have used $|\bm E_i^\prime|_+^2-|\bm E_i^\prime|_-^2=0$.

Through simple algebra, one has:
\begin{equation}\label{s28}
\langle \bm E^{\prime\ast}(\bm r)\cdot\nabla \phi_{\text{dipole}}(\bm r)\rangle_{\Omega+}-\langle \bm E^{\prime\ast}(\bm r)\cdot\nabla \phi_{\text{dipole}}(\bm r)\rangle_{\Omega-}=\frac{2\sqrt{\epsilon_r}}{3c}i|E_0|^2\frac{\bm m_{21}\cdot [\hat{K}^\dagger(\bm r)\nabla(\bm \Phi^{\text{tot}}(\bm r)\cdot \bm\mu_{21})]}{\hbar(\omega-\omega_0)+i \gamma_{21}-G}
\end{equation}
where we have used the enhancement factor $\hat{K}(\bm r)$ [Eq.\eqref{s36}]: $\bm E^\prime(\bm r)=\hat{K}(\bm r)\bm E_0=\bm e_0\hat{K}(\bm r)E_0$ with $\mathbf{E}^\prime(\bm r)$ being the filed induced only by NP dimer in the position $\bm r$ inside spheres. 

Substituting \eqref{s28} into \eqref{s26}, we have:
\begin{equation}\label{s29}
\text{CD}_{\text{NP, dipole-field}}=\frac{2\omega\sqrt{\epsilon_r}}{3c\pi\epsilon_0}|E_0|^2\sum_i(\mathbf{Im} \epsilon_{\text{NP}})\mathbf{Im}\int \frac{\bm m_{21}\cdot [\hat{K}^\dagger(\bm r)\nabla(\bm \Phi^{\text{tot}}(\bm r)\cdot \bm\mu_{21})]}{\hbar(\omega-\omega_0)+i \gamma_{21}-G} dV
\end{equation}

Next, we calculate the second term of $\text{CD}_{NP}$: $\text{CD}_{\text{dipole-dipole}}$. From Eqs. \eqref{s6} and \eqref{s7} we can see $\rho_{22}\propto |\sigma_{21}|^2$, so that $\text{CD}_{\text{mol}\pm}\propto \langle |\sigma_{21}|^2\rangle_{\Omega+}-\langle |\sigma_{21}|^2\rangle_{\Omega-}$. From Eq.\eqref{s22}, we have:
\begin{equation}\label{s30}
\langle |\sigma_{21}|^2\rangle_{\Omega+}-\langle |\sigma_{21}|^2\rangle_{\Omega-}=\frac{4\sqrt{\epsilon_r}}{3c}\frac{|E_0|^2\mathbf{Im}[\bm m_{21}\cdot(\hat{\mathcal{P}}^\dagger\bm \mu_{12})]}{|\hbar(\omega-\omega_0)+i \gamma_{21}-G|^2}
\end{equation}
thus:
\begin{align}\label{s31}
\text{CD}_{\text{NP, dipole-dipole}}&=\frac{\omega}{8\pi^2 \epsilon_0^2}\sum_i(\mathbf{Im}\epsilon_{\text{NP}})\left(\langle |\sigma_{21}|^2\rangle_{\Omega+}-\langle |\sigma_{21}|^2\rangle_{\Omega-}\right)\int |\nabla (\bm \Phi^{\text{tot}}\cdot \bm \mu_{21})|^2 dV\nonumber\\
&=\frac{\omega\sqrt{\epsilon_r}}{6\pi^2 \epsilon_0^2c}\frac{|E_0|^2\mathbf{Im}[\bm m_{21}\cdot(\hat{\mathcal{P}}^\dagger\bm \mu_{12})]}{|\hbar(\omega-\omega_0)+i \gamma_{21}-G|^2}J(\omega)
\end{align}
where we have defined: $J(\omega)=\sum_i(\mathbf{Im}\epsilon_{\text{NP}})\int |\nabla (\bm \Phi^{\text{tot}}\cdot \bm \mu_{21})|^2 dV$, where $-\frac{1}{4\pi \epsilon_0}\nabla (\bm \Phi^{\text{tot}}\cdot \bm \mu_{21})$ can be seen as the field induced by an effective molecular dipole $\bm \mu_{12}$. Thus, $J(\omega)$ has the meaning of energy rate.  Specifically, in the presence of molecule, the absorption rate of metal antiparticles due to energy transfer from molecule to spheres are given by: $Q=\frac{2\omega}{16\pi^2 \epsilon_0^2}\sum_i(\mathbf{Im}\epsilon_{\text{NP}})\int |\nabla (\bm \Phi^{\text{tot}}(\bm r)\cdot \bm\mu_{12})|^2dV$, meanwhile the rate of energy transfer of molecule is given by: $E=\omega\cdot 2\mathbf{Im}(-G)=-2\omega\mathbf{Im}G$. If we neglect the radiation from the system, thus the energy conservation is satisfied $Q=E$, resulting that: $J(\omega)=-16\pi^2 \epsilon_0^2\mathbf{Im}G$. Therefore, Eq.\eqref{s31} reduces to:
\begin{equation}\label{s32}
\text{CD}_{\text{dipole-dipole}}=-\frac{8\omega\sqrt{\epsilon_r}}{3c}\frac{|E_0|^2\mathbf{Im}[\bm m_{21}\cdot(\hat{\mathcal{P}}^\dagger\bm \mu_{12})]}{|\hbar(\omega-\omega_0)+i \gamma_{21}-G|^2}\mathbf{Im}G
\end{equation}

Once $\text{CD}_{\text{NP}}$ is obtained [\eqref{s29} and \eqref{s32}], the CD signal in the standard unit of $M^{-1}cm^{-1}$ can be obtained similarly to the derivation of $\text{CD}_{\text{mol}}$:
\begin{equation}\label{s33}
\Delta \sigma_{\text{NP}}=\frac{\text{CD}_{\text{dipole-field}}+\text{CD}_{\text{dipole-dipole}}}{\langle S\rangle_t},\text{CD}_{\text{NP}}: \Delta \varepsilon_{\text{NP}}=\frac{N_A}{0.23}\Delta \sigma_{\text{NP}}
\end{equation}

\section{Comparison of components of CD signal}
\begin{figure}[htbp]
\includegraphics[width=.8\columnwidth]{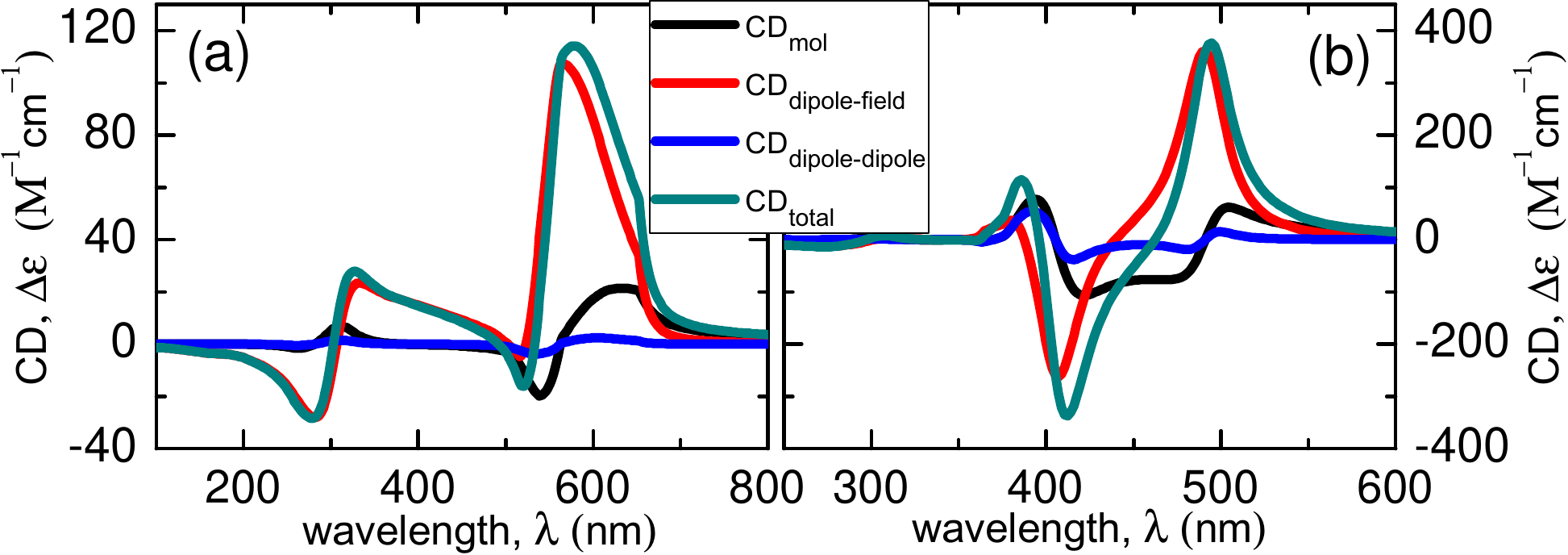}
\caption{ (color online) (a) shows the three components of CD signal with varying $\lambda$ under $\lambda_0=300nm$ and $d=1nm$ for Au dimer, while (b) is the same evolution of CD signal for Ag dimer. Molecule is placed in the center of dimer, i.e., $\Delta=R+d/2$, and other parameters are: $R=10nm, \bm \mu||z, \lambda_0=300nm$.}\label{SI1}
\end{figure}

Next, we see the comparison of components of CD signal. 

\begin{figure}[htbp]
\includegraphics[width=\columnwidth]{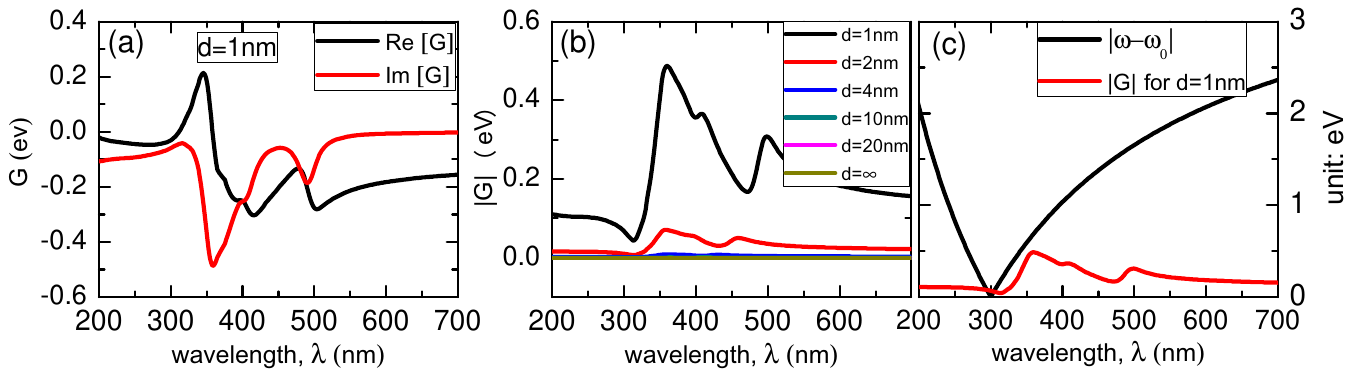}
\caption{ (color online). Graph (a) shows the real and imaginary parts of G function for $d = 1nm$ for a molecule-Ag dimer system in the case $\bm \mu\perp z$, while (b) gives the evolution of $|G|$ for diffident $d$ for the same system. (c) is the comparison of $|\omega-\omega_0|$ and $|G|$ under $d=1nm$. Other parameters are the same as in Fig.\ref{SI1}. }\label{SI2}
\end{figure}

In Fig.\ref{SI1} we show the three components of CD signal at $\lambda_0=300nm$ and small distance $d=1nm$ for gold NPs [(a)] and Ag NPs [(b)] respectively. In such off-resonant hybrid system,  the chiral molecule usually does not absorb at the plasmonic wavelength, but it provides a chiral non-absorbing environment for NPs because it is still optically active, resulting that the current inside NPs is chiral and accordingly the CD signal occurs. Considering that the NPs are not chiral, thus $\text{CD}_{\text{dipole-field}}$ and $\text{CD}_{\text{dipole-dipole}}$ are totally due to the exciton-plasmon interaction between molecule and NPs. We can see clearly a large plasmonic peak in CD spectrum emerges, while the $\text{CD}_{\text{dipole-field}}$ term makes the main contribution. In fact, in the presence of single NP rather than a dimer, $\text{CD}_{\text{mol}}$ and $\text{CD}_{\text{NP}}$ can be obtained analytically as\cite{ref1}: $\text{CD}_{\text{mol}}\propto 1/\Delta \omega^2, \text{CD}_{\text{NP}}\propto1/ \Delta \omega$ under large $\Delta \omega=\omega_0-\omega$ and $|\Delta \omega|\gg \gamma_{12}$. Besides, $\text{CD}_{\text{dipole-dipole}}\propto \mathbf{Im}G$ is usually very small in most cases unless for very small $d$, since $G\sim d^{-3}$. That's why $\text{CD}_{\text{dipole-field}}$ determines the CD signal at the plasmon frequency in this case. To see it clearly, in Fig.\ref{SI2} we show the order of $G$ function with varying separation $d$, from which we can see $G$ is usually very small in most cases ($|G|\ll \gamma_{21}, \omega_0$), and for not small separation, $d\geq 2nm$ for example, it can even be ignored. At very small separation, like $d=1nm$ in our simulation where the giant CD effect can be observed, although here $G$ is comparable with $\gamma_{21}$ ($\mathbf{Re}G\ll \omega_0, \mathbf{Im} G\sim \gamma_{21}$), the influence of $G$ function is also small at the plasmon frequency and $\text{CD}_{\text{dipole-field}}$ still dominates the main CD feature (see Fig.\ref{SI1}).

\section{Absorption rate of Ag dimer}

\begin{figure}[htbp]
\includegraphics[width=.8\columnwidth]{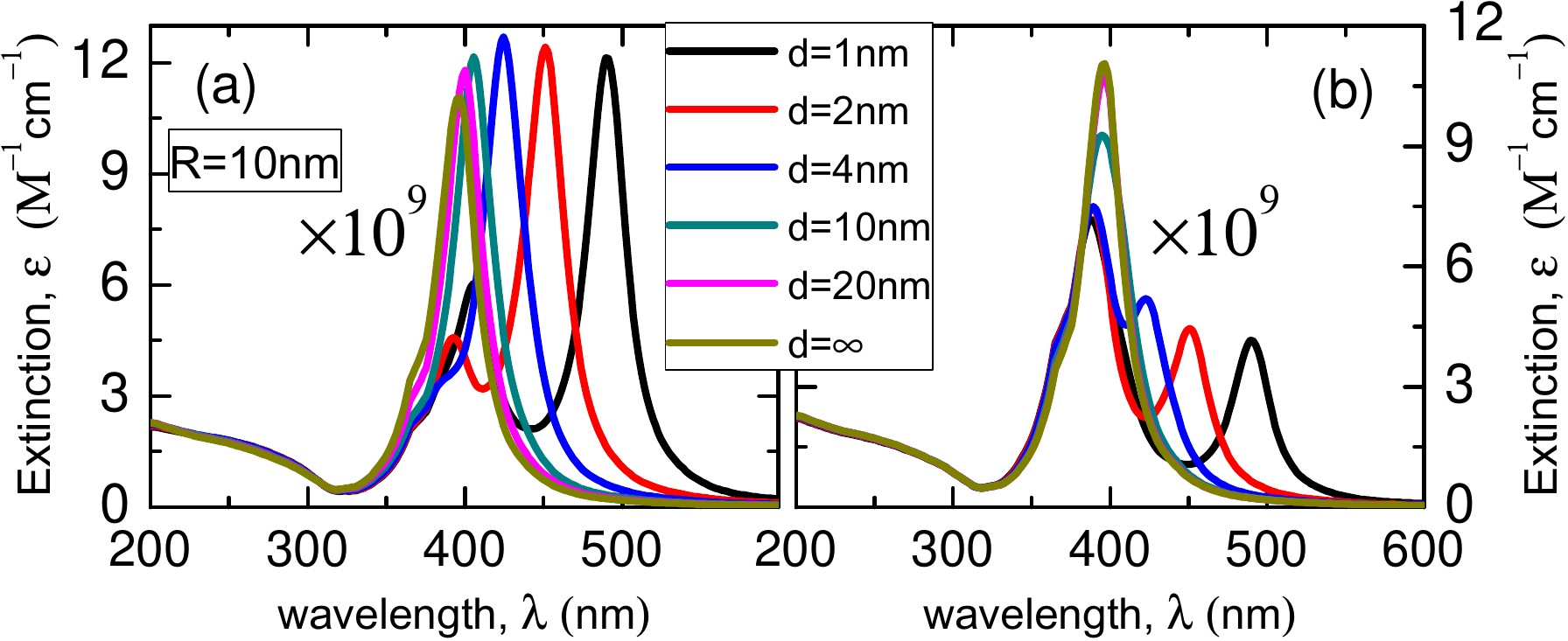}
\caption{ (color online) (a) is extinction of single Ag dimer v.s. $\lambda$ with varying $d$ under $\mathbf{E}_0||z$, while (b) is extinction of Ag dimer after averaging over the
molecular dipole orientation. Other parameters are the same as in Fig.\ref{SI1}.}\label{SI3}
\end{figure}

In Fig.\ref{SI3} we show the extinction of Ag dimer with varying distance $d$, and found that the height of multipole plasmonic peaks can be higher than the dipole peak. Specifically, without the average process $\langle\ldots\rangle_\Omega$, multipole plasmonic peaks emerge with lower height under small $d$ [see Fig.\ref{SI3}(a)]. However, since molecule-dimer complex actually has random orientations in experiments, the height of multipole peaks become surprisingly higher after averaging process [see Fig.\ref{SI3}(b)]. This is because the dipole peak will nearly vanish at $\mathbf{E}_0\perp z$ due to the inter-NPs Coulomb interaction, but the weight in averaging is the most biggest. Similar results can be obtained for Au dimer. 

\section{Molecular quadrupole contribution}

We now examine the effect of quadrupole light-matter interaction. We will show that the effect of quadrupole is negligible for the function $G$, but can be essential for the CD signals under certain conditions. The quadrupole term in the light-matter interaction operator $\hat{H}^\prime$ is:
\begin{equation}
V_{\text{quad}}=-\frac{e}{3}\sum_{\alpha,\beta}\frac{1}{2}\left(3r_\alpha r_\beta-r^2 \delta_{\alpha,\beta}\right)\frac{\partial E_\alpha}{\partial r_\beta}=-\frac{1}{3}\sum_{\alpha,\beta}Q_{\alpha,\beta}\frac{\partial E_\alpha}{\partial r_\beta}
\end{equation}
where $E_\alpha$ is the $\alpha$ component of the total electric field inside the molecule $\mathbf{E}_{T}$.

For an exact calculation of $G_{\text{quad}}$, one can employ the same procedure as above for the dipolar term $G_\text{dipole}$. In the case of quadrupole, $\mathbf{d}\rightarrow Q_{\alpha,\beta}$ and $\mathbf{\Phi}\rightarrow \Psi^{\alpha,\beta}(\alpha, \beta=x, y, z)$. For example, in the presence of molecular quadrupole $Q_{\alpha, \beta}=\frac{1}{2}\int\left(3 r_\alpha r_\beta-r^2 \delta_{\alpha \beta}\right) \rho(\mathbf{r})dV$, the electric field coming from the charges on surface of NPs induced only by quadrupole is given by:
\begin{equation}
\mathbf{E}_{\text{quad-NP},\omega}(\mathbf{r})=-\frac{1}{4\pi \epsilon_0}\nabla\sum_{\alpha,\beta}Q_{\alpha\beta,\omega}\left[\Psi_1^{\alpha,\beta}(\mathbf{r})+\Psi_2^{\alpha,\beta}(\mathbf{r})\right]
\end{equation}
where $\Psi_i^{\alpha,\beta}$ is an electric potential induced by the $i$th NP in the presence of quadrupole: $Q_{\alpha\beta, \omega}: \varphi_{\text{quad},i}=\frac{1}{4\pi \epsilon_0}\sum_{\alpha \beta}Q_{\alpha,\beta}\Psi_i^{\alpha,\beta}(\mathbf{r})$. 
Then, the total $G$ function will be:
\begin{align}
G&=\frac{1}{4\pi \epsilon_0}\bm \mu_{21}\cdot\nabla\left.\left[(\mathbf{\Phi}_1^{\text{out}}+\mathbf
{\Phi}_2^{\text{out}})\cdot \bm \mu_{12}+\sum_{\alpha^\prime, \beta^\prime}Q_{\alpha^\prime \beta^\prime,12}(\Psi_1^{\alpha^\prime \beta^\prime}+\Psi_2^{\alpha^\prime \beta^\prime})\right]\right|_{\mathbf{r}_0}\nonumber\\
&\hspace{5em}+\frac{1}{4\pi \epsilon_0}\frac{1}{3}\sum_{\alpha,\beta}Q_{\alpha \beta,21}\frac{\partial}{\partial r_\alpha}\frac{\partial}{\partial r_\beta}\left.\left[(\mathbf{\Phi}_1^{\text{out}}+\mathbf
{\Phi}_2^{\text{out}})\cdot \bm \mu_{12}+\sum_{\alpha^\prime, \beta^\prime}Q_{\alpha^\prime \beta^\prime,12}(\Psi_1^{\alpha^\prime \beta^\prime}+\Psi_2^{\alpha^\prime \beta^\prime})\right]\right|_{\mathbf{r}_0}\nonumber\\
&\equiv G_{\text{dipole}}+G_{\text{quad}}+G_{\text{interaction}}
\end{align}
where:
\begin{align}
&G_{\text{dipole}}=\frac{1}{4\pi \epsilon_0}\bm \mu_{21}\cdot\nabla(\mathbf{\Phi}_1^{\text{out}}+\mathbf
{\Phi}_2^{\text{out}})\cdot \bm \mu_{12}\\
&G_{\text{quad}}=\frac{1}{4\pi \epsilon_0}\frac{1}{3}\sum_{\alpha,\beta}Q_{\alpha \beta,21}\frac{\partial}{\partial r_\alpha}\frac{\partial}{\partial r_\beta}\left[\sum_{\alpha^\prime, \beta^\prime}Q_{\alpha^\prime \beta^\prime,12}(\Psi_1^{\alpha^\prime \beta^\prime}+\Psi_2^{\alpha^\prime \beta^\prime})\right],
\end{align}
and the term $G_{\text{interaction}}$ describes a coupling between the dipolar and quadrupole transitions induced by the NP-induced fields. Noticeably, if the molecule is located at the symmetry point of system, which is the case in our model, then the interaction term $G_{\text{interaction}}$ vanishes.

Then, by using the expansion coefficients of quadrupole in terms of spherical harmonics and also boundary condition, the final result of $G$ becomes:
\begin{align}
&G=G_{\text{dipole}}+G_{\text{quad}}\\
&G_{\text{dipole}}=\frac{1}{4\pi \epsilon_0}\sum_{lm}\left[\left(\sum_i \mu_{21,i}A_{lm}^{i\ast}\right)\frac{\sum_{j}C_{lm}^j \mu_{12,j}}{r_{d,1}^{l+2}}+\left(\sum_i \mu_{21,i}F_{lm}^{i\ast}\right)\frac{\sum_{j}D_{lm}^j \mu_{12,j}}{r_{d,2}^{l+2}}\right]\\
&G_{\text{quad}}=\frac{1}{4\pi \epsilon_0}\frac{1}{3}\sum_{lm}\left[\left(\sum_{\alpha \beta}Q_{\alpha \beta,21}\mathbb{A}_{lm}^{\alpha \beta\ast}\right)\frac{\sum_{\alpha^\prime \beta^\prime}
\mathbb{C}_{lm}^{\alpha^\prime \beta^\prime}Q_{\alpha^\prime \beta^\prime,12}}{r_{d,1}^{l+3}}+\left(\sum_{\alpha \beta}Q_{\alpha \beta,21}\mathbb{F}_{lm}^{\alpha \beta\ast}\right)\frac{\sum_{\alpha^\prime \beta^\prime}
\mathbb{D}_{lm}^{\alpha^\prime \beta^\prime}Q_{\alpha^\prime \beta^\prime,12}}{r_{d,2}^{l+3}}\right]
\end{align}
Here $\mathbb{A}_{lm}^{i,j}(\mathbb{F}_{lm}^{ij})$ are expansion coefficients similar to $A_{lm}^i(F_{lm}^i)$, while coefficients $\mathbb{C}_{lm}^{i,j}(\mathbb{D}_{lm}^{ij})$ are obtained through boundary conditions.

\begin{figure}[htbp]
\includegraphics[width=.8\columnwidth]{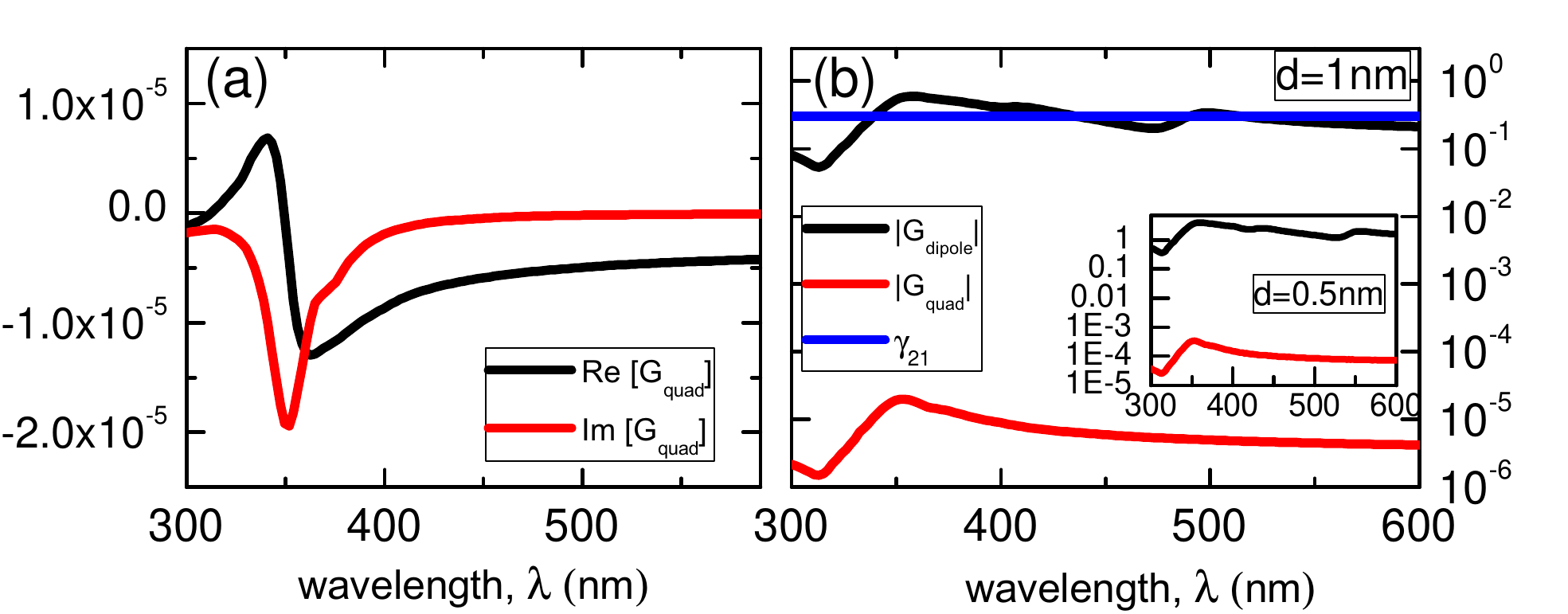}
\caption{ (color online) (a) The real and imaginary parts of $G_{\text{quad}}$ for $d=1nm$ for a molecule-Ag dimer system in the case $\bm \mu\perp z$. (b) The functions $|G_{\text{dipole}}|, |G_{\text{quad}}|$ for $d=1nm$ for the same system; the inset shows the same for $d=0.5nm$. Ohter parameters are: $R=10nm, Q_{xx,21}=e(0.25\AA)^2, Q_{yy,21}=e(0.25\AA)^2/4, Q_{zz,21}=e(0.25\AA)^2, Q_{xy,21}=Q_{yx,21}=e(0.25\AA)^2/2, Q_{xz,21}=Q_{zx,21}=e(0.25\AA)^2,Q_{yz,21}=Q_{zy,21}=e(0.25\AA)^2/2$.}\label{SI4}
\end{figure}

We now look at the numerical results for the quadrupole contributions to the optical spectra. For the quadrupole components, we will take numbers typical for molecules\cite{Barron-paper}: $\sqrt{Q_{\alpha \beta,21}/e}=r_{\text{mol}}\sim 0.25\AA$. Then, taking $G$ function, for example, we see that the quadrupole contribution, $|G_{\text{quad}}|$, is really small compared to the dipolar term. In Fig.\ref{SI4}, we show the comparison of $|G_{\text{dipole}}|, |G_{\text{quad}}|$ and also $\gamma_{21}$. From this figure we can see even at very small separation $d=1nm (0.5nm)$, the influence of molecular quadrupole on $G$ function can be neglected. This can also be understood by looking at the equations for $|G_{\text{dipole}}|$ and $|G_{\text{quad}}|$. Since $r_{\text{mol}}\sim r_{12}$,
\begin{equation*}
\frac{G_{\text{quad}}}{G_{\text{dipole}}}\sim\frac{r_{\text{mol}}^2}{D^2},
\end{equation*}
where $D$ is a characteristic size of the system that should be taken here as $d$. Then we see that $G_{\text{quad}}\ll G_{\text{dipole}}$ since $r_{\text{mol}}\ll d$ even for the smallest separation $d\sim 0.5nm$.

Now we look at the contributions of quadrupole interaction to the CD signals. The formalism develops as in the case of the electric and magnetic dipoles. After averaging over the solid angle of incident light and collecting the leading terms (the needed terms should be $\sim kD$, where $D$ is a characteristic dimension of the system), we have: 
\begin{align}
&\text{CD}_{\text{mol,quad}}=2 \omega_0\frac{\gamma_{12}}{|\hbar(\omega-\omega_0)+i \gamma_{12}-G|^2}\frac{2k}{9}|E_0|^2\times\nonumber\\
&\hspace{4em}2\mathbf{Re}i\left.\left(\mu_{21,x}\mathcal{P}_{xx}Q_{yz,21}\left[\frac{\partial^2 \phi_{e_y,k_z}^\ast}{\partial r_y\partial r_z}-\frac{\partial^2 \phi_{e_z,k_y}^\ast}{\partial r_y\partial r_z}\right]+\mu_{21,y}\mathcal{P}_{yy}Q_{zx,21}\left[\frac{\partial^2 \phi_{e_z,k_x}^\ast}{\partial r_z\partial r_x}-\frac{\partial^2 \phi_{e_x,k_z}^\ast}{\partial r_z\partial r_x}\right]\right)\right|_{\mathbf{r}_0=0}
\end{align}
where the function $\phi_{e_\alpha,k_\beta}$ originate from the retarded effects in the response of the NP dimer on the external electromagnetic field and are defined as:
\begin{equation*}
\nabla\cdot \epsilon(\mathbf{r})\nabla \phi_{e_\alpha,k_\beta}=i\frac{\partial}{\partial r_\alpha}\left(\epsilon(\mathbf{r})r_\beta\right)
\end{equation*}
These function can be found numerically using the expansion over spherical harmonics related to individual NPs. The plasmonic contribution to the quadrupole CD is:
\begin{align}
&\text{CD}_{\text{NP,quad-field}}=-\mathbf{Im}(\epsilon_{\text{NP}})|E_0|^2\frac{\omega}{2\pi \epsilon_0}\frac{2k}{9}\int_{V_{\text{NP}}}dV\mathbf{Re}\frac{2i}{\hbar(\omega-\omega_0+i \gamma_{12}-G)}\times\nonumber\\
&\left[
\begin{aligned}
&\left\{\left(K_{xx}\right)^\ast\frac{\partial \Phi_x^{\text{tot}}}{\partial x}+\left(K_{yx}\right)^\ast\frac{\partial \Phi_x^{\text{tot}}}{\partial y}+\left(K_{zx}\right)^\ast\frac{\partial \Phi_x^{\text{tot}}}{\partial z}\right\}\times\left(\mu_{21,x}Q_{yz,21}\right)\left[\frac{\partial^2 \phi_{e_y,k_z}}{\partial y\partial z}-\frac{\partial^2 \phi_{e_z,k_y}}{\partial y\partial z}\right]\\
&+\left\{\left(K_{xy}\right)^\ast\frac{\partial \Phi_y^{\text{tot}}}{\partial x}+\left(K_{yy}\right)^\ast\frac{\partial \Phi_y^{\text{tot}}}{\partial y}+\left(K_{zy}\right)^\ast\frac{\partial \Phi_y^{\text{tot}}}{\partial z}\right\}\times\left(\mu_{21,y}Q_{zx,21}\right)\left[\frac{\partial^2 \phi_{e_z,k_x}}{\partial x\partial z}-\frac{\partial^2 \phi_{e_x,k_z}}{\partial x\partial z}\right]
\end{aligned}\right]
\end{align}
where the response  function $\mathbf{\Phi}^{\text{tot}}$ comes from the molecular dipole so that the total field inside the NPs induced by the incident wave and by the molecule is written as:
\begin{equation}
\mathbf{E}_{\omega,\text{inside NP}}=\mathbf{E}_\omega^\prime-\frac{1}{4\pi \epsilon_0}\nabla\left(\mathbf{\Phi}_\omega^{\text{tot}}\cdot\mathbf{d}_\omega\right)-\frac{1}{4\pi \epsilon_0}\nabla\sum_{\alpha \beta}Q_{\alpha\beta,\omega}\Psi_\omega^{\alpha \beta,\text{tot}}(\mathbf{r})
\end{equation}
Again, the function $\Phi_\omega^{\text{tot}}$ should be calculated numerically.

\begin{figure}[htbp]
\includegraphics[width=.6\columnwidth]{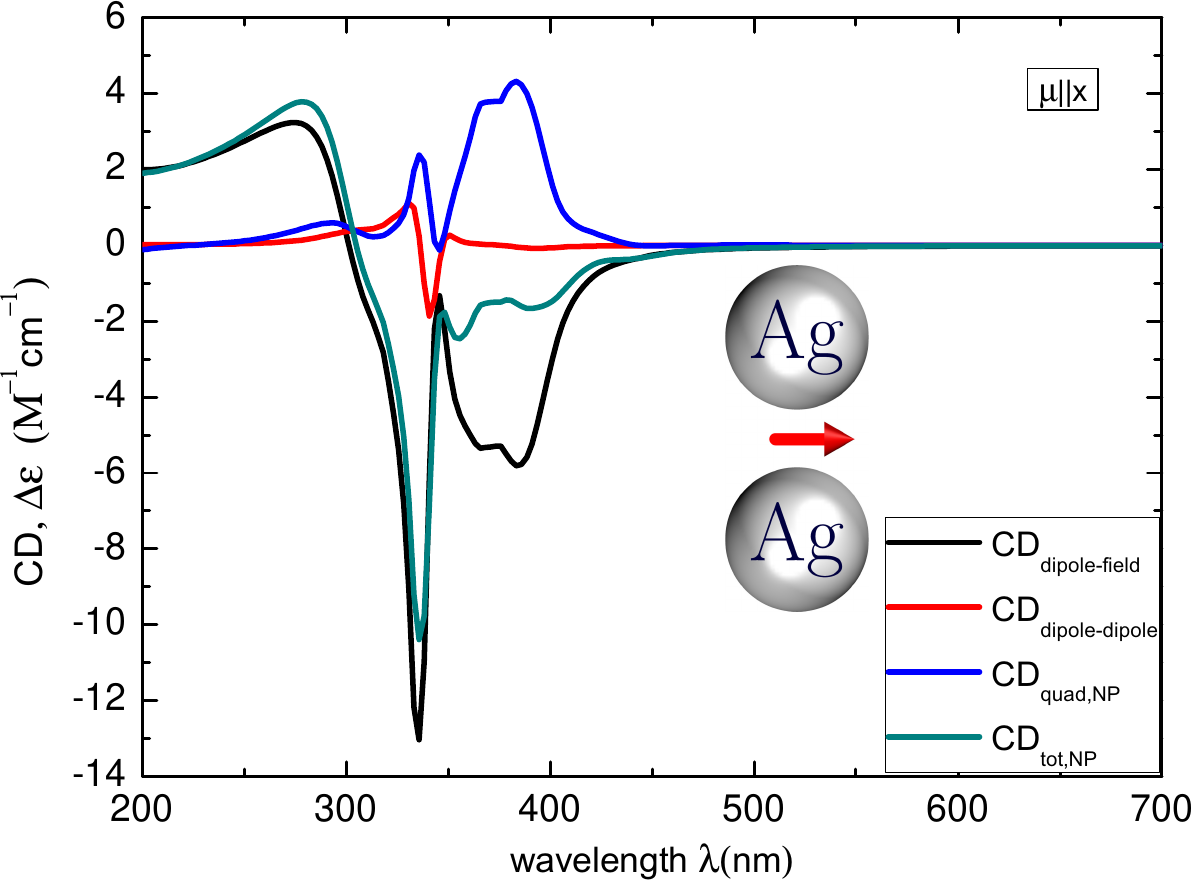}
\caption{ (color online) CD spectra of Ag dimer with $R=10nm, d=1nm$ and $\bm \mu_{12}||x$. On can see that the quadrupole CD makes an essential contribution to the total CD signal. Unmentioned molecular parameters are the same as in Fig.\ref{SI4}.}\label{SI5}
\end{figure}

\begin{figure}[htbp]
\includegraphics[width=.6\columnwidth]{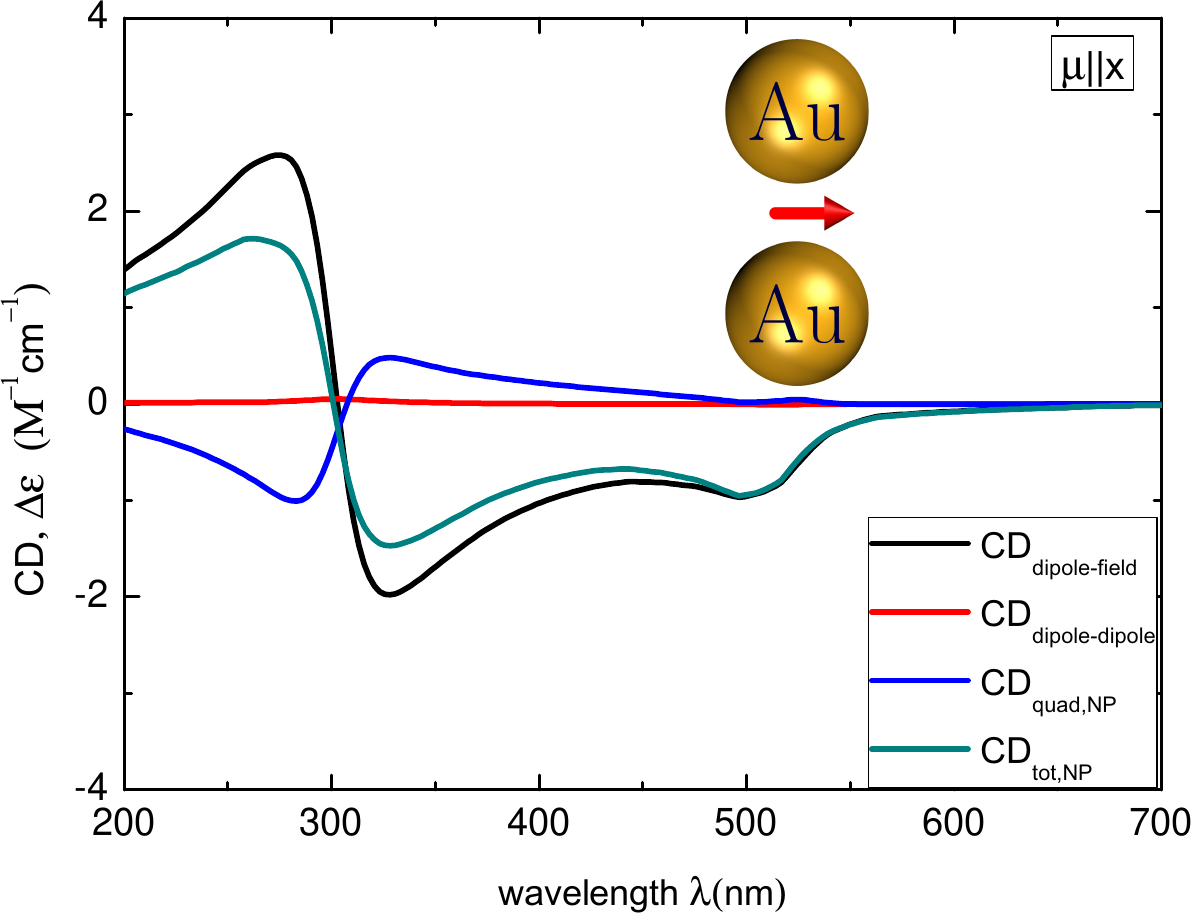}
\caption{ (color online) CD spectra of Au dimer with $R=10nm, d=1nm$ and $\bm \mu_{12}||x$. On can see that the appearance of an essential quadrupole contribution. Unmentioned molecular parameters are the same as in Fig.\ref{SI4}.}\label{SI6}
\end{figure}

Numerical results for the quadrupole CD with a molecule $\bm \mu_{12}||x$ are shown in Fig.\ref{SI5} and Fig.\ref{SI6}. Importantly, because of the symmetry of the system, the quadrupole CD depends only on $\mu_{21,x}$ and $\mu_{21,y}$ and does not depend on $\mu_{21,z}$. Therefore, the quadrupole CD is extended not to be strongly enhanced since the plasmonic enhancement in the hot spot occurs only for the electric field parallel to the $z$ direction and therefore only $\mu_{21,z}$ is involved in the enhancement. Simultaneously, the terms in the CD signal related the $\mu_{21,x(y)}$ may be even suppressed because of the dynamic screening. Therefore, the main result for the plasmon-enhanced CD for the case $\bm \mu_{12}||z$ (Figs. (2b) and (3a) in the main text) are not affected by the quadrupole term, whereas the CD spectra of a molecule with $\bm \mu_{12}||x(y)$ can be modified by the quadrupole effect (Figs.\ref{SI5} and \ref{SI6} below and Fig.2c in the main text). 

Another important model is randomly-oriented molecules inserted into the plasmonic hot spots in an ensemble of NP dimers. In this case, we can use the above equations for the functions $\text{CD}_{\text{quad}}$ and averaged them over all ordinations of a molecule. Of course, such averaging gives zeros CD: $\langle \text{CD}_{\text{quad}}\rangle_{\text{molecular orientation}}=0$ since $\langle Q_{xz(yz)}\rangle_{\text{rotation of molecule about $y(x)$}}=0$. Therefore, as can be expected from the very beginning, the quadrupole contribution to the CD signal of randomly-oriented chiral molecules inserted into symmetry hot spots of NP dimers should vanish. The reason is the high symmetry of the NP system.

Overall we can see that, due to the high symmetry of the plasmonic antenna, the quadrupole contributions are not essential for the effect of giant plasmonic CD. This is somewhat expected since the contribution of the quadrupole interaction to CD of molecules in an isotropic medium is averaged to zero, and the quadrupole CD in an isotropic matrix becomes important only for oriented molecules\cite{Barron-paper,Barron-book}. In our case, we deal with randomly-oriented dimers, but the symmetry for the molecule becomes lowered (due to the presence of NP dimer), and therefore, the quadrupole CD can be somewhat essential under the peculiar conditions $\bm \mu_{12}||x(y)$. But, the central effect of giant plasmonic CD remains relatively insensitive to the quadrupole transitions.

